%% file: Phase.tex
\documentclass[12pt,a4paper,final]{iopart}

\usepackage{iopams}  
\usepackage{graphicx}
\usepackage[breaklinks=true,colorlinks=true,linkcolor=blue,urlcolor=blue,citecolor=blue]{hyperref}

\begin{document}

\title[The phase structure of QCD]{The phase structure of QCD}

\author{Christian Schmidt}
\address{Fakult\"at f\"ur Physik, Universit\"at Bielefeld, D-33615 Bielefeld, Germany}
\ead{schmidt@physik.uni-bielefeld.de}

\author{Sayantan Sharma}
\address{Physics Department, Brookhaven National Laboratory, Bldg. 510A, Upton, NY 11973, USA}
\ead{sayantans@quark.phy.bnl.gov}

\begin{abstract}
We review recent results on the phase structure of QCD and bulk QCD
thermodynamics.  In particular we discuss how universal critical
scaling related to spontaneous breaking of the chiral symmetry
manifests itself in recent lattice QCD simulations and how the
knowledge on non-universal scaling parameter can be utilized in the
exploration of the QCD phase diagram. We also show how various
(generalized) susceptibilities can be employed to
characterize properties of QCD matter at low and hight temperatures,
related to deconfining aspects of the QCD transition. Finally, we
address the input these lattice QCD calculation can provide for our
under standing of the matter created in heavy ion collisions and in
particular on the freeze-out conditions met in the hydrodynamic evolution 
of this matter.
\end{abstract}

\pacs{12.38.Mh, 21.65.Qr, 25.75.Nq}
\vspace{2pc}

\section{Introduction}
The theory that describes the fundamental interaction between quarks
and gluons is called Quantum Chromodynamics (QCD).
Ever since its formulation people have speculated on the QCD phase diagram.
QCD is constructed as a non
abelian gauge theory and posses two most astonishing features:
asymptotic freedom \cite{Politzer:1973fx, Gross:1973id} and a global
symmetry (chiral symmetry) that is spontaneously broken at small
temperatures and net baryon number densities
\cite{Nambu:1961tp}. These are the principles that not only guide our thinking
of the spectrum of QCD bound states in vacuum, but give also rise to a
rich and fascinating phase structure of QCD.  
Even before the fundamental properties of QCD had been fully revealed, 
it was
realized that the spectrum of hadronic resonances that becomes
exponentially dense at high energies, would naturally lead to the
concept of a limiting temperature in a gas of hadrons
\cite{Hagedorn:1965st}.  Later on, with the analysis of the running
QCD coupling that due to asymptotic freedom vanishes in the high
energy limit, it was quickly understood that hadronic matter undergoes
a transition to a new state of matter, the quark gluon plasma (QGP)
\cite{Cabibbo:1975ig, Collins:1974ky}.  At low temperatures and net
baryon number densities the coupling constant is so large, that quarks
are tightly bound into hadrons.  The static quark potential is in fact
linear rising with distance, thus quarks are confined into
hadrons.  However, once the density of hadrons increases coursed by
thermal fluctuations or a large baryon chemical potential, they start
to overlap and the potential gets screened. Hence, the quarks get
liberated and are allowed to move over macroscopic distances \cite
{Baym:1979, Celik:1980td}.

\begin{figure}
\begin{center}
\includegraphics[width=0.425\textwidth, height=0.4\textwidth]{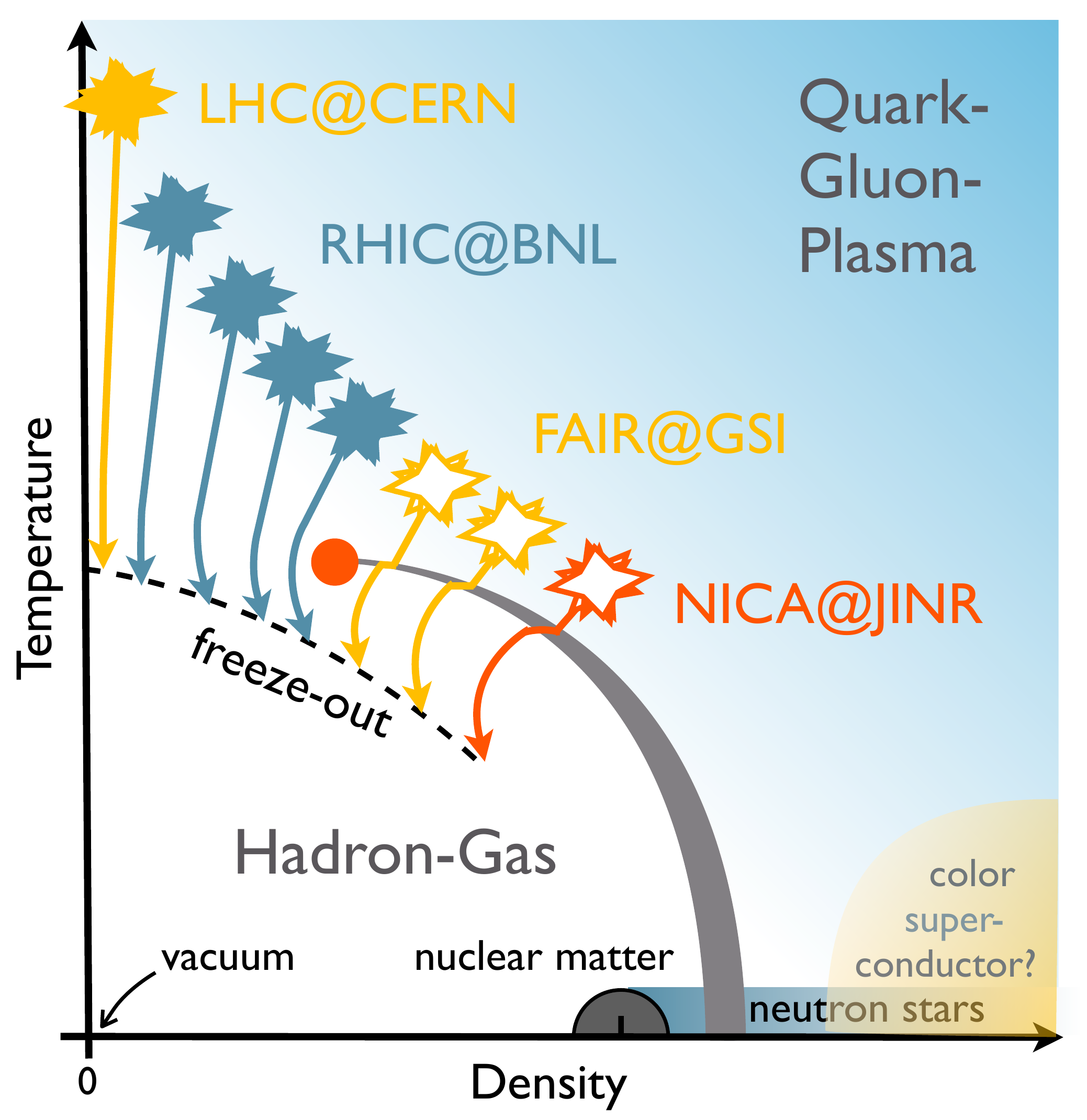}
\end{center}
\caption{Schematic phase diagram of QCD. The red circle indicates a
  possible critical point, as an endpoint of a line of first order
  phase transitions that separate the hadron gas phase from the quark
  gluon plasma phase.  Also shown are approximate initial conditions
  of the fireballs that are created in recent and future heavy ion
  collisions, as well as a sketch of their hydrodynamical evolution
  trajectories upon chemical freeze-out.
\label{fig:pdiag}}
\end{figure}

It was also speculated that the deconfinement transition of QCD goes
along with the global restoration of chiral symmetry, that is manifest
in the QCD Lagrangian \cite{Pisarski:1983ms}.  Based on the symmetries
of QCD many effective models have been designed and analyzed at non
zero temperature and baryon number density, which greatly enhanced our
understanding of the QCD phase diagram.  From these model calculations
a complex phase structure with various homogeneous as well as
in-homogeneous phases has emerged \cite{Fukushima:2010bq}.  At very
high net-quark number densities, several different pairing mechanisms
among quarks have been proposed, that would lead to various color
super-conducting phases, similar to electrical super-conductors in
condensed matter systems \cite{Rajagopal:2000wf,
  Alford:2001dt}. Unfortunately, details of those phases are highly
model dependent and could so far not be addressed with first principle
QCD calculations. For this reason we show in Fig.~\ref{fig:pdiag} a
simplified version of the QCD phase diagram that mainly features a
speculative critical point, that appears as a critical end-point (CEP)
of a line of first order transitions. This CEP is currently subject to
many lattice QCD calculations as well as experimental heavy ion
programs.

The framework of lattice regularized QCD provides a non perturbative
method, which has proven to be very successful since its inception
\cite{Wilson:1974sk, Creutz:1980zw}. As the QCD coupling constant is
rather large at temperatures close to the QCD transition
\cite{Prosperi:2006hx} and since it is running logarithmically, it
becomes small only at exponentially high energy scales. This is the
reason why ordinary perturbation theory is expected to be reliable
only at very high temperatures and net baryon number densities.  Using
appropriate re-summation schemes, as hard thermal loop (HTL)
perturbation theory \cite{Andersen:1999fw} or electric QCD (EQCD)
\cite{Braaten:1995jr} the range of validity can be extended down to a
few times the transition temperature ($T_c$). However, close to $T_c$
QCD is deeply non perturbative. Here lattice QCD is the only known
method that -- after a proper continuum extrapolation -- can address
the thermodynamics of quarks and gluons.  This is definitely true for the case of a
vanishing net baryon number, or equivalently, a vanishing baryon
chemical potential ($\mu_B$). At non zero chemical potential, lattice
QCD suffers from the infamous sign problem. At small $\mu_B$, however,
this problem can be overcome by a Taylor expansion in $\mu_B/T$
\cite{Allton:2002zi}. Other methods that have the potential to solve
the sign problem completely are currently being developed
\cite{Aarts:2015kea, Gattringer:2016kco}.
 
At the same time, huge experimental effort is undertaken to create and
measure the thermal properties of QGP. Shortly after the formation of
man-made QGP matter in a heavy ion experiment was first announced at
the Super Proton Synchrotron (SPS) at CERN \cite{Heinz:2000bk}, the
Relativistic Heavy Ion Collider (RHIC) at BNL went in operation. At
RHIC the envisaged goal was exactly to crate and study QGP.  Also at
the Large Hadron Collider (LHC) a dedicated heavy ion program is
conducted.  Related to the dynamical expansion of the QGP medium,
after its creation in the central collision region of two heavy
nuclei, RHIC discovered a few very remarkable transport properties of
QGP.  Based on the observation of elliptic flow and jet quenching
effects, it is believed that in a short temperature range directly
above $T_c$, the QGP behaves as an almost perfect fluid, with a small
shear viscosity and small transport coefficients such as electrical
conductivity and diffusion coefficients.  It is tempting to study
those transport properties also on the lattice, which will provide
crucial input to the modeling of the expansion of the QGP.

More recently, RHIC initiated a beam energy scan program (BES) aiming
on the exploration of the QCD phase diagram and the discovery of the
postulated CEP.  Reducing the bombarding energy of the collision,
leads to a variation of the initial conditions at which the QGP is
created.  Lower bombarding energies translate into colder but more
dense systems at formation times.  With the planed future facilities
FAIR and NICA, one wants to increase the density even further.
Approximate initial conditions of various heavy ion experiments are
sketched in Fig.~\ref{fig:pdiag}.  Event-by-event fluctuation
observables are supposed to be sensitive to the phase structure of QCD
\cite{Stephanov:1999zu}.  Among these the moments of conserved charge
distributions are very interesting observables as they can also be
obtained from lattice QCD calculations. From a comparison of results, 
as obtained on the lattice and from the experiment, one can infer state 
variables of the system at the time of freeze-out \cite{Bazavov:2012vg}.

In this article we review recent selected (lattice) QCD results on QCD
bulk thermodynamics that are relevant for heavy ion phenomenology.
After a short presentation of the basic lattice methods, we start with
the equation of state in Sec.~\ref{sec:eos}.  We will then discuss
universal critical behavior, that can be observed in the lattice
results close to second order phase transition points.  We emphasize
how non-universal constants that match the lattice data to the
universal scaling functions can be utilized to extract information on
the phase diagram in Sec.~\ref{sec:crit}. One important class of
observables in which critical behavior can be made manifest are the
before mentioned cumulants of charge fluctuations. These can, however,
also be used to demonstrate the liberation of quarks from the hadronic
bound states for $T>T_c$, which is done in Sec.~\ref{sec:dof}. 
And finally in Sec.~\ref{sec:freeze}, we discuss
several phenomenological results that can be obtained from a
comparison of experimental cumulants of conserved charge fluctuations
with the corresponding lattice results.

\section{QCD bulk thermodynamics on the lattice\label{sec:eos}}
\subsection{The path integral on the lattice}
The fundamental quantity to calculate 
thermal expectation values is the partition function. As we are dealing with
a quantum field theory where particle pair creation and annihilation is at the 
heart of things, we will perform our simulations in the grand canonical
ensemble. The partition function $Z$ is obtained in the path integral formalism
as functional integral over all fundamental fields, weighted by the exponential 
of the negative Euclidean action functional. It is important to notice that thermal 
averages are calculated in Euclidean space, which is obtained after Wick rotation 
$t\to-it=\tau$. The temperature is introduced through a compactification of the 
$\tau$ direction, in analogy to classical statistical physics. The Euclidean space-time is 
discretized, thus coordinates can take values on a four dimensional hyper-cubic 
lattice,  $x_\nu = an_\nu$, with $n_\nu\in\{1,2,\dots, N_\nu\}$, where 
$\nu\in\{\tau,x,y,z\}$ and $a$ denotes the dimensionful lattice constant. 
Here we restrict ourselves to isentropic lattices where $a$ is not depending on 
the direction $\nu$. Furthermore, we consider cubic volumes with 
$N_x=N_y=N_z\equiv N_\sigma$ and periodic boundary conditions. For a 
given lattice $N_\tau\times N_\sigma^3$, temperature and spatial volume are 
obtained as
\begin{equation}
T=\frac{1}{aN_\tau}\qquad\mbox{and}\qquad V=(aN_\sigma)^3\;.
\label{eq:T}
\end{equation}

In lattice QCD, the fundamental fields  are the Grassmannian quark 
$\psi^{(f)}_n$ and anti-quark fields $\bar\psi^{(f)}_n$ for each quark flavor ($f$), 
as well as the link fields $U_{n,\nu}$ that are associated with the links between 
neighboring sits. They describe a parallel transport of the continuum gluon field 
$A_\nu(x)$ from site $x$ to $x+a\hat\nu$ 
\begin{equation}
U_{x,\nu}=P \exp\left\{i g \int_x^{x+a\hat\nu}{\rm d}x_\nu A_\nu(x)\right\}\;,
\end{equation}
where $\hat\nu$ is the unit vector in $\nu$-direction and $P$ denotes path 
ordering. The Link fields take values in the $SU(3)$ color group. Similarly, 
the Dirac 4-spinors $\psi^{(f)}_n$ and $\bar\psi^{(f)}_n$ are color 3-vectors. 
The spinor and color indices have been suppressed here. The number of the 
quark flavors $N_f$, as well as the corresponding quark masses $m_f$, 
with $f\in\{1,2,\dots,N_f\}$ are parameters of QCD. Here we will -- if not stated 
differently -- discuss (2+1)-flavor QCD, {\it i.e.} we will have two flavor of mass 
degenerate light quark ($u,d)$, with masses $m_u=m_d$ and one heavier strange
quark $s$, with mass $m_s$. We also introduce one quark chemical potentials 
for each quark flavor $\vec\mu=(\mu_u,\mu_d,\mu_s)$. Finally, we arrive at an 
expression for the partition function, which is given as 
\begin{equation}
Z(T,V,\vec\mu)=\int\prod_{n,\nu}{\rm d}U_{n,\nu}
\int\prod_{n,f}{\rm d}\psi^{(f)}_n\;{\rm d}\bar\psi^{(f)}_n \;e^{-S_E(T,V,\vec\mu)}\;,
\label{eq:Z1}
\end{equation}
where the measure ${\rm d}U_{n,\nu}$ denotes the Haar measure on the 
compact gauge group $SU(3)$. Note that on the lattice the path integral is not a 
functional integral any more but a mathematically well defined multi-dimensional 
integral.

The Euclidean action $S_E$ is composed of a fermionic and a gluonic part $S_E=S_F+S_G$. 
The gluon action is to leading order in $a$ given by Wilson's plaquette action \cite{Wilson:1974sk}, 
where a plaquette variable $U_{n,\nu\rho}$ is defined as the product of link variables along 
an elementary plaquette 
$U_{n,\nu\rho}\equiv U_{n,\nu}U_{n+\hat\nu,\rho}U^\dagger_{n+\hat\rho,\nu}U^\dagger_{n,\rho}$. 
With this definition the gluon action is given as
\begin{equation}
S_G=\sum_n\sum_{0\le \nu<\rho\le 3}\beta\left(1-\frac{1}{3}{\rm Re}{\rm Tr}U_{n,\nu\rho}\right)\;.
\label{eq:SG}
\end{equation} 
The fermionic part has the structure of a quadratic form 
\begin{equation}
S_F=\sum_f\sum_{n,m}\bar\psi^{(f)}_nM^{(f)}_{n,m}\psi^{(f)}_m\;,
\end{equation}
with a fermion matrix $M^{(f)}_{n,m}$. Different forms of this matrix define
different fermion discretizations on the lattice. Without giving a specific form of the 
fermion matrix, we will discuss properties of different lattice fermions in Sec.~\ref{sec:ferm}. 

The Grassmannian quark fields  $\psi^{(f)}_n$ and $\bar\psi^{(f)}_n$ may be integrated 
out analytically from Eq.~(\ref{eq:Z1}), which is most desirable since handling 
Grassmannians on the computer is impractical. The partition function may then be 
written as
\begin{equation}
Z(T,V,\vec\mu)=
\int\prod_{n,\nu}{\rm d}U_{n,\nu}\prod_{f} {\rm det}\left[M^{(f)}(\mu_f)\right] \;e^{-S_G(T,V)}\;,
\label{eq:Z2}
\end{equation}
featuring the fermion determinant ${\rm det}\left[M^{(f)}(\mu_f)\right]$. 
Here we have made explicit that the chemical potential enters only in the fermion 
part of the action, {\it i.e.} in the fermion matrix. It is introduced by a modification of 
the temporal links as $U_{n,0}\to e^{a\mu_f}U_{n,0}$ and $U^\dagger_{n,0}\to e^{-a\mu_f}U^\dagger_{n,0}$ 
\cite{Hasenfratz:1983ba}. Other possibilities to introduce the chemical potential  have 
also been considered \cite{Gavai:1985ie, Gavai:2014lia}.

The connection to bulk thermodynamics is done by standard thermodynamic relations 
among the thermodynamic potentials. Here we employ the grand canonical potential 
$\Omega(T,V,\vec\mu)\equiv -T{\rm ln} Z(T,V,\vec\mu)$ from which the pressure can 
be obtained most easily as $p(T,V,\vec\mu)=-\Omega(T,V,\vec\mu)/V$. However, any 
other thermal expectation value $\left<\mathcal{O}\right>(T,V,\vec\mu)$ of an 
observable $\mathcal{O}$ can be obtained by 
\begin{equation}
 \left<\mathcal{O}\right>=\frac{1}{Z}
\int\prod_{n,\nu}{\rm d}U_{n,\nu}\;\mathcal{O} \prod_{f} {\rm det}\left[M^{(f)}(\mu_f)\right] \;e^{-S_G(T,V)}\;.
\end{equation}
In fact, since the partition function itself is hardly accessible from MC simulations, it is common practice
to study derivatives of $\ln Z$, taken with respect to various external parameter. Such derivatives
are then obtained through the calculation of thermal expectation values, e.g. we have 
$\partial \ln Z / \partial m_{u} =<Tr[(M^{(u)})^{-1}]>$. This strategy is also followed for the calculation 
of the energy density and pressure (see Sec.~\ref{sec:trace_anomaly}). 
Equally important to infer on the properties of the QGP 
is the study of correlation-functions, resulting in the 
evaluation of observables of the type $<\mathcal{O}_x \mathcal{O}_y>$. Although 
seemingly very difficult, it is possible to reconstruct spectral-functions from
those Euclidean correlators by either Bayesian or variational methods. The review of 
of correlation- and spectral-functions are, however, beyond the scope of this article.

\subsection{Lattice discretization schemes and the continuum limit\label{sec:ferm}}
The discretization that has been performed in order to perform the MC Integration 
Eq.~(\ref{eq:Z2}) bears certain difficulties. The most obvious is certainly the introduction 
of discretization errors of the action functionals.  The gauge action as given in 
Eq.~(\ref{eq:SG}) receives corrections which are of order $\mathcal{O}(a^2)$.
These discretization errors can be systematically removed by adding additional terms to the  
gauge action that are composed of Wilson loops of length greater than that of the plaquette 
\cite{Symanzik:1983dc, Symanzik:1983gh}. In Fact, it is common to add in a suitable linear 
combination the sum of all Wilson loops of rectangular shape. Such a Symanzik improved 
gauge action is correct up to correction of order $\mathcal{O}(a^4)$.

With in the fermionic sector the problems are more severe. Here the naive nearest-neighbor 
discretization of the covariant derivative in the Dirac operator not only introduces discretization
errors of order  $\mathcal{O}(a)$ it also gives rise to unphysical degrees of freedom. This is 
due to the fact that a finite lattice spacing introduces a cut-off in momentum space and thus a periodicity 
in the lattice dispersion relation. The fermion modes that are associated with the roots of the dispersion
relation in the $2^d$ corners of the Brillouin zone are known as fermion doublers. Moreover, 
the fermion doublers are intimately connected with  the explicit breaking of the chiral symmetry 
on the lattice \cite{Nielsen:1981hk}. Several discretization schemes have been invented to
deal with these problems. Among them are Wilson \cite{Wilson:1974sk}  and staggered 
\cite{Kogut:1974ag} fermions which are most commonly used due to their moderate computational costs, 
but also Domain Wall \cite{Kaplan:1992bt} and Overlap \cite{Neuberger:1997fp, Neuberger:1998wv} fermions.

In lattice calculations at nonzero temperature, the consequences of
the chosen fermion discretization scheme are three-fold. For a study
of universal critical behavior that arises due to the spontaneous
breaking of the chiral symmetry of QCD, it is most important to use an
action that possess at least a remnant of this symmetry. The situation
here is the following: The Wilson action breaks the chiral symmetry
completely at any finite lattice spacing. The staggered action
preserves a $U(1)_{\rm even}\times U(1)_{\rm odd}$ subgroup of the
symmetry, {\it i.e.} the fermion fields on even and odd sides can be
multiplied separately by a phase.  Domain-Wall fermions are chiral in
the limit of a large fifth dimension and Overlap fermions posses an exact (albeit modified)
chiral symmetry at any finite lattice spacing.  The latter two action
are so costly that so far no scaling study close to the chiral limit
has been possible. First studies of the transition temperature 
\cite{Bhattacharya:2014ara}, thermodynamic quantities \cite{Borsanyi:2012xf, Borsanyi:2015zva} 
and the anomalously broken axial symmetry $U_{\rm A}(1)$ and the Dirac spectrum \cite{Bazavov:2012qja,
  Buchoff:2013nra, Cossu:2013uua, Chiu:2013wwa, Sharma:2016cmz, Tomiya:2016jwr} have, however, been performed recently.

For studies of bulk thermodynamic quantities such as the pressure $p$ and the energy density $\epsilon$ the high and the low
temperature regimes suffer from two different types of discretization errors. At very high temperatures, 
deviations from the Stefan-Boltzmann (SB) limit stem from the discretization errors of the covariant derivative. Just as in the 
gluonic case, such errors can be eliminated systematically order by order using so called improved actions, which are 
widely used in case of Wilson and staggered fermions. For Wilson fermions, the clover-term \cite{Sheikholeslami:1985ij}
reduces the discretization errors of the action to $\mathcal{O}(a^2)$, for staggered fermions, straight (Naik \cite{Naik:1986bn}) or 
bended (p4   \cite{Heller:1999xz}) three-link terms diminish the error to $\mathcal{O}(a^4)$. It is interesting to note that even 
though the standard Wilson action features cutoff errors of $\mathcal{O}(a)$, corrections to bulk thermodynamic quantities are 
universal to standard Wilson and standard staggered actions and start at $\mathcal{O}(a^2)$. To summarize, an expansion 
of the dimensionless quantity $p/T^4$ around the ideal gas limit yields \cite{Hegde:2008nx}
\begin{equation}
\frac{\frac{p}{T^4}}{(\frac{p}{T^4})_{\rm SB}}=1+\left\{
\!\begin{array}{rcrcrl}
\frac{248}{147}(\frac{\pi}{N_\tau})^2&\!\!+\!\!&\frac{635}{147}(\frac{\pi}{N_\tau})^4&\!\!+\!\!&\frac{13351}{8316}(\frac{\pi}{N_\tau})^6&\mbox{std. Wilson}\\
\frac{248}{147}(\frac{\pi}{N_\tau})^2&\!\!+\!\!&\frac{635}{147}(\frac{\pi}{N_\tau})^4&\!\!+\!\!&\frac{3796}{189}(\frac{\pi}{N_\tau})^6&\mbox{std. staggered}\\
&\!\!-\!\!&\frac{1143}{980}(\frac{\pi}{N_\tau})^4&\!\!-\!\!&\frac{365}{77}(\frac{\pi}{N_\tau})^6&\mbox{Naik}\\
&\!\!-\!\!&\frac{1143}{980}(\frac{\pi}{N_\tau})^4&\!\!+\!\!&\frac{73}{2079}(\frac{\pi}{N_\tau})^6&\mbox{p4}
\end{array}\right. .
\end{equation}

At low temperatures, bulk thermodynamic quantities are, as we will
see, mainly determined by the hadronic spectrum of QCD.  Due to the
existence of fermion doublers on the lattice -- also called fermion
tastes -- the hadronic states are duplicated as well.  Moreover, high
energy gluons that scatter fermions from one corner of the Brillouin
zone to another, introduce a nontrivial interaction between different
fermion tastes, which leads to the so called taste splitting of
hadronic states. This splitting is most prominent in the light pion
sector. Within the staggered fermions formulations, the doublers are
reduced from 16 to 4 tastes. Consequently, with these 4 tastes many
more (15) pions can be formed, which become degenerated only in the
continuum limit. In order to suppress large momentum modes of the
gluons and thus reduce the taste splitting, smeared gauge fields are
used in the Dirac operator.  Popular staggered type actions are the
2stout and 4stout actions, that use 2 and 4 levels of stout-smearing
\cite{Morningstar:2003gk}, respectively as well as the HISQ (highly
improved staggered quark) action \cite{Follana:2006rc}. The latter
combines two levels of 5- and 7-link smearing with the Naik-term that
also improves the high-T limit of this action. In terms of taste
splitting, which can be measures, {\it e.g.}, by the root-mean-square
mass $M_\pi^{\rm RMS}$ of the pions, it is found that the 4stout
actions performs similarly well as the HISQ action.

Having understood the leading correction in the lattice spacing $a$,
we eventually extrapolate to the continuum limit ($a=0$) by using
calculations performed at several lattice spacings and fitting
appropriate polynomials to it. This extrapolation is done at fixed
volume $V$ and temperature $T$. The lattice spacing is controlled by
the lattice coupling $\beta=6/g^2$. Given the relation
Eq.~(\ref{eq:T}), for a lattice with a fixed number of points in
temporal direction $N_\tau$, a temperature scan is performed by
varying $\beta$.  At small values of the bare coupling $g$ (large
$\beta$) the relation between $\beta$ and $a$ and thus $\beta$ and $T$
is given by the well known perturbative $\beta$-function of QCD. In
practice, at temperatures close to the transition, the coupling is,
however, not small and the scale has to be set by using an
experimentally measured input observable, which is compared to the
same quantity measured on the lattice, in units of the lattice spacing
$a$. As this input observable is almost always a hadronic quantity,
experimentally determined in vacuum, the scale setting is performed at
$T=0$, {\it i.e.} on lattices with $N_\tau\le N_\sigma$. In principle
one is completely free to chose the input observable, as in the
continuum limit all scale ambiguities should vanish. For recent
thermodynamic calculations, quantities related to the static quark
potential ($r_0$ \cite{Sommer:1993ce}, $r_1$ \cite{Bernard:2000gd}),
the kaon decay constant ($f_K$) or quantities related to the Wilson
flow ($t_0^{1/2}$ \cite{Luscher:2010iy}, $w_0$ \cite{Borsanyi:2012zs})
have been used for scale setting.  It is worth noting that the
continuum extrapolation is done with fixed renormalized quark
masses. This requires a non trivial tuning of the bare quark masses as
function of $\beta$. At first, the strange quark mass $m_s$ is fixed
to its physical value by demanding that the mass of the $\eta_{s\bar
  s}$ meson, which is related to the pion and kaon by chiral
perturbation theory, is given by $M_{\eta_{s\bar
    s}}=\sqrt{(2M_K^2-M_\pi^2)}=686$ MeV. 
Having fixed $m_s$, we fix
the light quark mass by holding the ration $m_l/m_s$ constant. 
Some of the results we will present 
have been obtained with $m_l=ms/20$, although the physical value is $m_l=ms/27$.
The relation between the coupling $\beta$ and the bare quark masses is
called line of constant physics (LCP).
 
\subsection{The equation of state\label{sec:trace_anomaly}}
\begin{figure}
\begin{center}
\includegraphics[width=0.325\textwidth, height=0.325\textwidth]{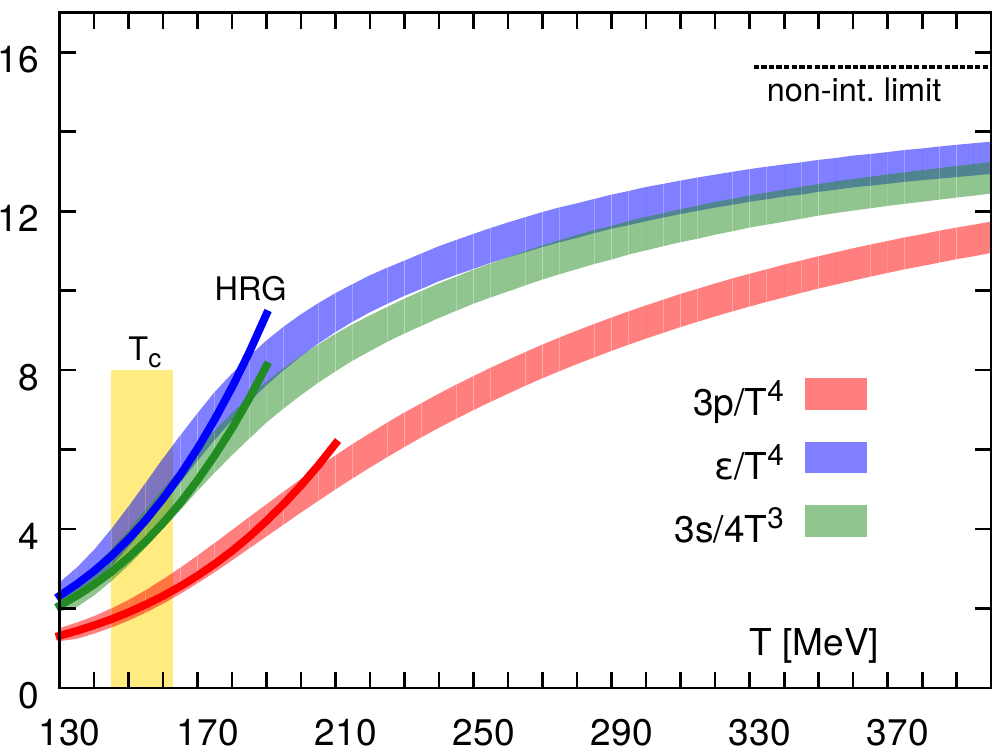}
\includegraphics[width=0.325\textwidth, height=0.325\textwidth]{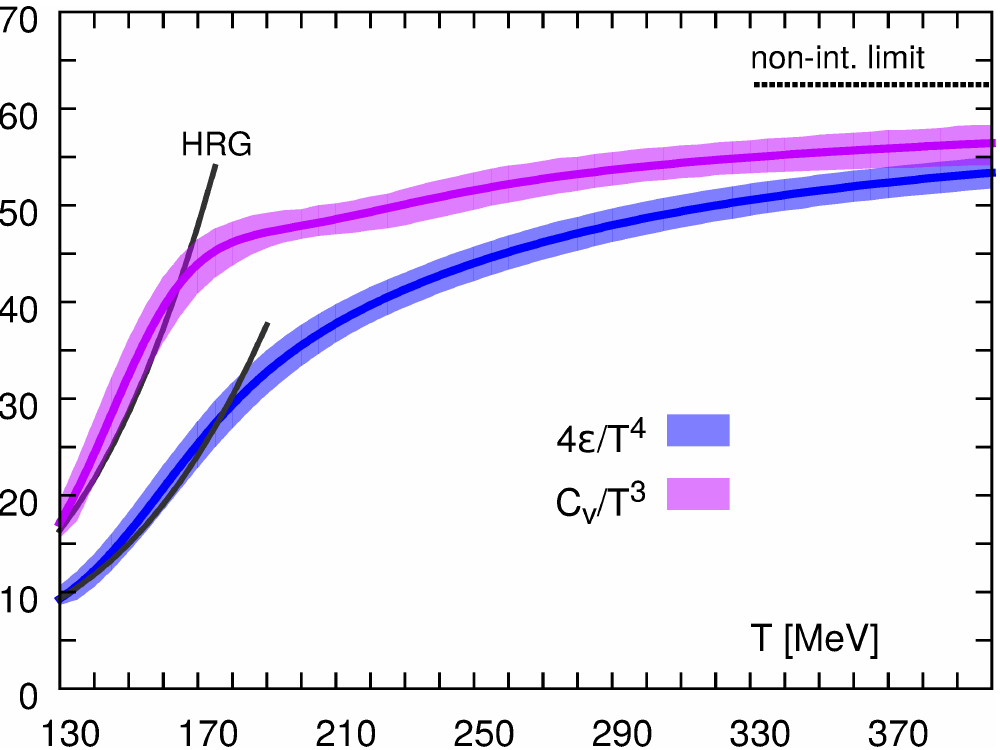}
\includegraphics[width=0.325\textwidth, height=0.325\textwidth]{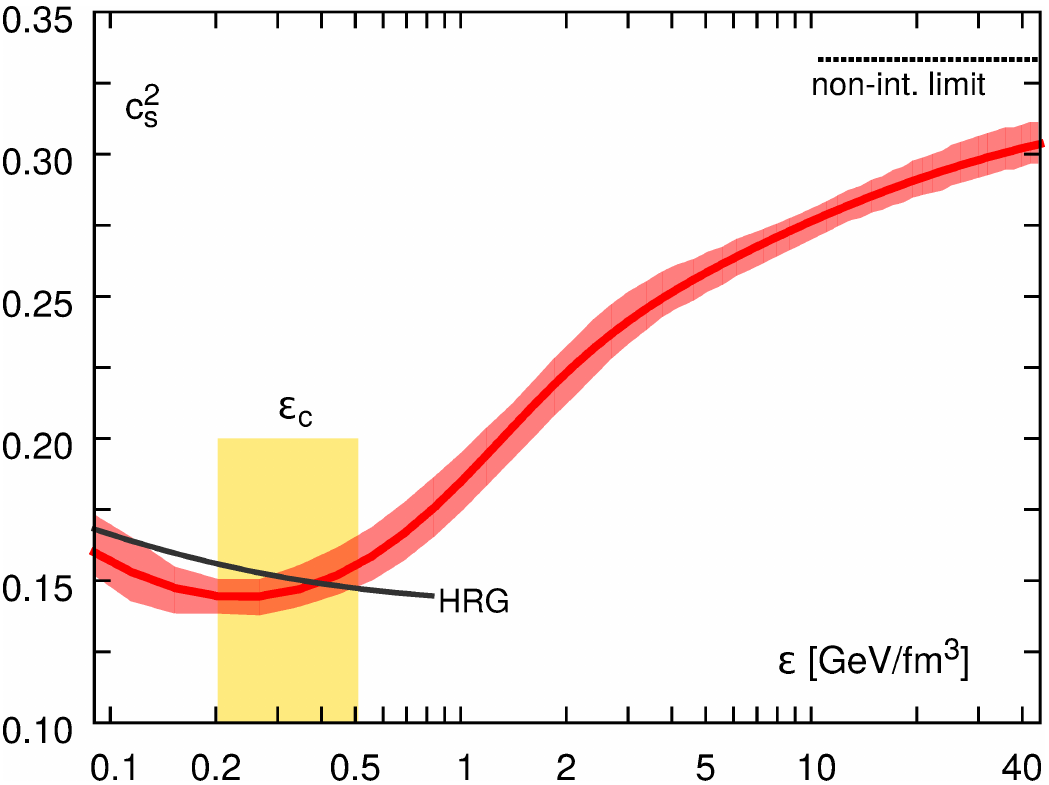}
\end{center}
\caption{Continuum extrapolated and suitably normalized pressure,
  energy density, and entropy density as a function of the temperature
  (left), the energy density and specific heat (middle) and the speed
  of sound squared versus energy density (right) from
  Ref.~\cite{HotQCDeos}.  A 2\% scale error is included in the error
  bands.  Dark lines indicate the HRG prediction and the horizontal
  lines correspond to the ideal gas limit. The vertical band marks the
  crossover region, $T_c=(154\pm 9)$ MeV on the left panel and the
  corresponding range in energy density, $\epsilon_c=(0.18-0.5)$
  $\mbox{GeV/fm}^3$ on the right panel.
\label{fig:hotqcd_eos}}
\end{figure}

The trace of the energy-momentum tensor,  also called trace anomaly or
the interaction measure ($\Theta^{\mu\mu}$), is related to the pressure $p$ by a temperature derivative (see, {\it e.g.} Ref.~\cite{Cheng:2007jq})
but can also be written as a total derivative of $\ln Z$ with respect to the lattice spacing
\begin{equation}
\frac{\Theta^{\mu\mu}(T)}{T^4} =\frac{\epsilon -3p}{T^4}=T\frac{d}{d T}\left(\frac{p}{T^4}\right) =-\frac{1}{T^3V}\frac{d\ln Z}{d \ln a}\; ,
\label{theta_p}
\end{equation}
where $\epsilon$ is the energy density. It is straight forward to evaluate the right-hand side of the last equality on the lattice, 
thus $\Theta^{\mu\mu}(T)$ is the fundamental observable. The pressure $p$ can be calculated by the integral method, {\it i.e.} by 
integrating $\Theta^{\mu\mu}(T)/T^5$, up to an integration constant.
This also allows to reconstruct the energy density $\epsilon$ as well as the entropy density $s/T^3=(\epsilon + p )/T^4$.
Recently continuum extrapolated results have been reported \cite{Borsanyi:2013bia, HotQCDeos}, using stout and HISQ actions, respectively.
The calculations are in good agreement, although the light quark masses are slightly different ($m_l=ms/27$ \cite{Borsanyi:2013bia} vs. $m_l=ms/20$ \cite{HotQCDeos}).
In Fig.~\ref{fig:hotqcd_eos} we show the HISQ results \cite{HotQCDeos}.

The rise of the pressure, energy density, and entropy density close to the transition region of 
$T_c=159\pm9$ MeV \cite{HotQCD_Tc} is a consequence of the liberation of many new quark degrees of freedom.
It is interesting to note that the critical energy density, associated with the transition region $\epsilon_c=0.34\pm 0.16$
$\mbox{GeV/fm}^3$  is  only slightly larger than the critical energy density of ordinary nuclear matter
$\epsilon^{\mbox{nuclear matter}}\approx 0.15$ $\mbox{GeV/fm}^3$. In fact, $\epsilon_c$ is close to the energy density
reached in the dense packing limit of nucleons, assuming a nucleon radius of $R_N=0.8$ $\mbox{fm}$.
   
Also shown in Fig.~\ref{fig:hotqcd_eos} are dark solid lines that indicate the hadron resonance gas results at low temperature.
The hadron resonance gas is a non interacting gas of all known hadronic bound states and resonances. It is eminent from Fig.~\ref{fig:hotqcd_eos}
that at low temperatures the QCD results come very close the HRG. A more detailed comparison of the HRG and QCD results of the pressure, is given 
in section Sec.~\ref{sec:dof}. Indeed it is found that a departure of the QCD pressure, and thus a fade of hadronic degrees of freedom starts just after the 
transition regions. On the other hand, the bulk thermodynamic quantities obtained from HISQ fermions \cite{HotQCDeos} 
comes close to the hard thermal loop perturbative calculation \cite{Andersen:2015eoa}
and the dimensional reduced QCD (EQCD) calculation \cite{Laine:2006cp} at temperatures $T\gtrsim 400$ $\mbox{MeV}$.
At such temperatures all three calculations differ on the 10\% level (for a comparison see Ref. \cite{HotQCDeos}). 
We conclude that the temperature region $T_c \le 2 T_c$ is deconfined but highly 
nonperturbative, {\it i.e.} the QGP is strongly coupled in this temperature range.

Of high interest is also the specific heat, $C_V={\rm d}\epsilon/{\rm d}T$, and speed of sound square, $c_s^2={\rm d}p/{\rm d}\epsilon$, which are also
shown in Fig,~\ref{fig:hotqcd_eos}. In a hydrodynamic description of the fireball that is generated in an heavy ion collision, these quantities are, {\it e.g.}, 
directly related to the life time of the fireball, and thus to the question of whether the plasma has enough time to equilibrate or not. We find that the 
specific head does not show a peak although in the chiral limit ($m_l\to 0$) it should develop a cusp, {\it i.e.} critical behavior dictates 
$C_V\sim\left(\frac{T-T_c}{T_C}\right)^{-\alpha}+C_V^{\rm reg}$, with $\alpha\approx-0.2$. The regular term is thus dominating.

\section{Universal critical behavior and the QCD phase diagram\label{sec:crit}}
The chiral symmetry of the QCD Lagrangian plays an important role for the determination of the QCD phase structure. In the limit 
of vanishing quark masses this symmetry is $SU_L(N_f)\times SU_R(N_f)\times U_A(1)\times U_V(1)$, which implies that left 
handed and right handed spinors can be rotated separately. The latter $U_V(1)$ symmetry represents the baryon number 
conservation. The $U_A(1)$ is anomalously broken by quantum fluctuations. Nonzero quark masses break the 
$SU_L(N_f)\times SU_R(N_f)$ symmetry explicitly, however, more importantly it is also broken spontaneously at low
temperatures. This gives, {\it e.g.} rise to three (almost) massless pseudo-scalar Goldstone bosons, that are identified with the pions.

\begin{figure}
\begin{center}
\includegraphics[height=0.4\textwidth]{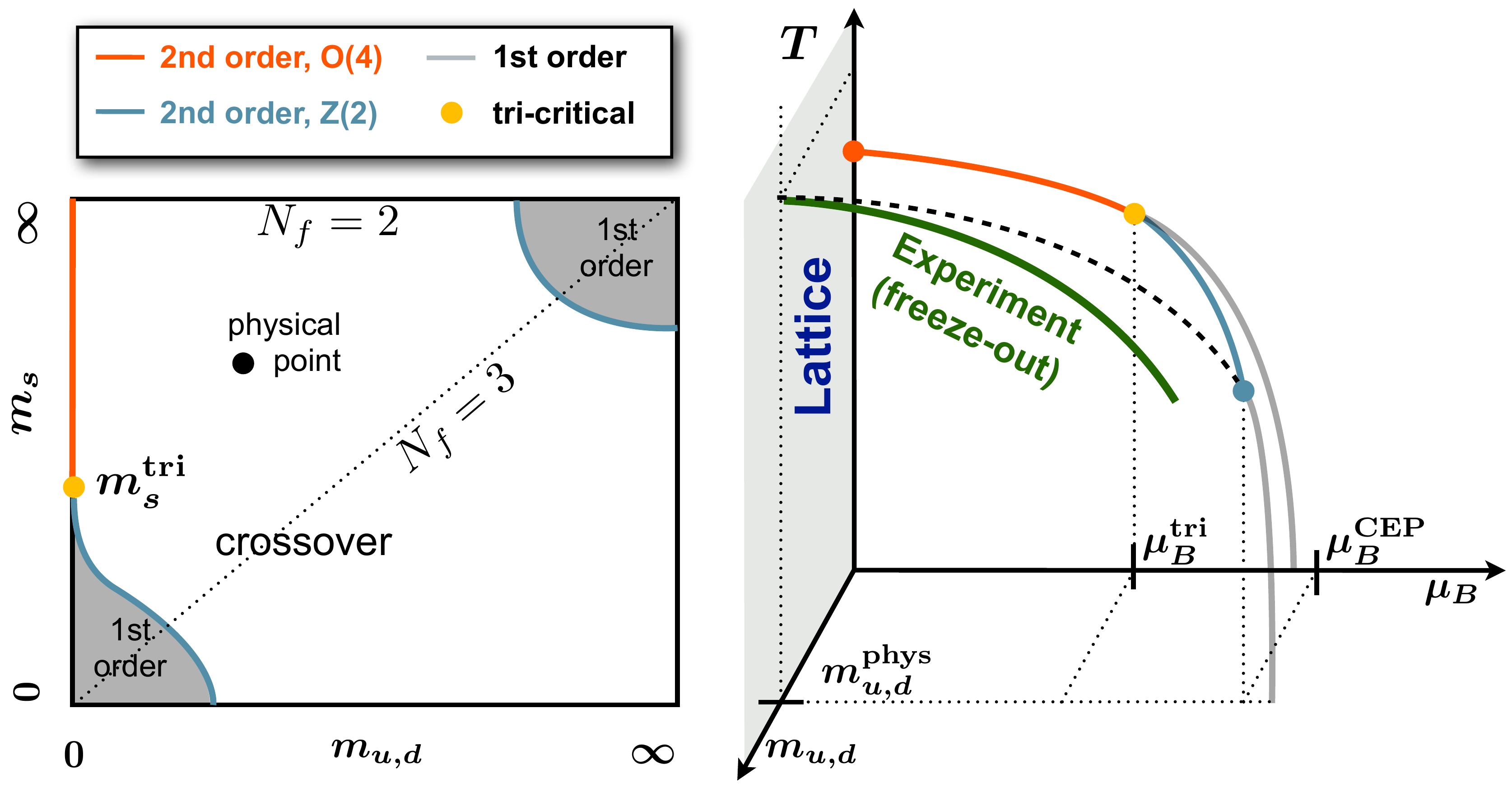}
\end{center}
\caption{A sketch of the nature of the QCD transition as functions of
  the two degenerate light (up and down) quarks with masses,
  $m_{u,d}=m_l$, and a heavier strange quark with mass, $m_s$, at zero
  baryon chemical potential (left) and the the expected QCD phase
  diagram in the ($T,\mu_B$)-plane for (2+1)-flavor QCD at fixed
  physical strange quark mass, in dependence of $m_{u,d}$
  (right). Also shown on the right are the regions of the phase
  diagram that are accessible to direct lattice calculations and to
  the experiment.
\label{fig:columbia}}
\end{figure}

The nature of the chiral transition crucially depends on the quark masses and the number of flavors. In 
Fig.~\ref{fig:columbia} (left), a sketch of the nature of the QCD phase diagram with two mass degenerate light flavors 
(up and down) and one heavier strange quark at $\mu_B=0$ is show. 
For the bulk part of the diagram, the QCD transition at $\mu_B=0$ is
not a true phase transition, but a rapid continuous and analytic
crossover. This includes also the physical mass point
\cite{Aoki:2006we}.  In the limit of infinite quark masses ($N_f=0$,
pure gauge theory), which is realized in the upper right corner of the
diagram we have a region of first order transitions. The same is true
for the lower left corner, the massless limit of three degenerate
quark masses ($N_f=3$). On the boundary of both of these regions the
transition becomes second order, with an universality class belonging
to the 3d-Z(2) (Ising) model. For the lower left corner this has been
demonstrated with std. staggered \cite{Karsch:2001nf,
  deForcrand:2003vyj} and Wilson-clover fermions
\cite{Takeda:2015bna,Kuramashi:2016kpb}. Simulations with improved
staggered fermions have so far not seen a first order region. Only an
upper limit for the degenerate quark mass $m_u=m_d=m_s<m_s^{\rm
  pyhs.}/270$ could be given, where $m_s^{\rm pyhs.}$
\cite{Endrodi:2007gc, Bazavov_3f, Ding:2011du} denotes the physical value of the
strange quark mass.

The situation in the chiral limit of two degenerate flavor is yet to
be clarified. It is dependent on the strength of the axial anomaly
($U_A(1)$ breaking) close to the transition temperature $T_c$. If the
$U_A(1)$ breaking is still significant at $T_c$, the transition is of
second order and since $SU(2)\times SU(2)$ is isomorphic to $O(4)$, it
is in the universality class of the 3dim.  $O(4)$ model
\cite{Pisarski:1983ms}. However, if the anomaly is already
sufficiently restored at $T_c$, the transition is either first or
second order with a different symmetry breaking pattern
\cite{Grahl:2013pba, Pelissetto:2013hqa}. For recent lattice studies
of the $U_A(1)$ anomaly see, {\it e.g.} \cite{Bazavov:2012qja,
  Buchoff:2013nra, Cossu:2013uua, Chiu:2013wwa, Sharma:2016cmz,
  Tomiya:2016jwr}.  The situation is expected to carry over to the
chiral limit of the (2+1)-flavor theory (holding $m_s$ fixed and
sending $m_{u,d}\to 0$), as long as the strange quark mass is lager
than a tri-critical strange quark mass $m_s^{\rm tri}$ where we enter
the first order region in the lower left corner of the
diagram. Unfortunately, the relative location of $m_s^{\rm tri}$ with
respect to the physical strange quark mass $m_s^{\rm pyhs.}$ is
unclear. It is thus not known whether $m_s^{\rm tri}<m_s^{\rm pyhs}$
or $m_s^{\rm tri}>m_s^{\rm pyhs}$ is realized in nature, which means
that in the chiral limit of a (2+1)-flavor theory with a fixed strange
quark mass we could either find a first or second order
transition. The former would give rise to a critical $Z(2)$ point at
finite quark masses. Even the possibility of $m_s^{\rm tri}\to \infty$
is not excluded. It was recently discussed and supported by lattice
simulations at imaginary chemical potential with std. staggered
fermions on coarse lattices \cite{Bonati:2014kpa}.  In the following
discussion we will, however, assume $m_s^{\rm tri}<m_s^{\rm pyhs.}$,
which is consistent with the universal scaling observed with improved
staggered fermions \cite{Ejiri:2009ac, Kaczmarek:2011zz,
  Bazavov:2011nk}.

The extension of this diagram to a nonzero baryon chemical potential
$\mu_B$ is also controversially discussed. Especially interesting is
whether the first order region in the lower left corner is shrinking or
growing with increasing $\mu_B$. One possible extension is shown in
the right-hand side of Fig.~\ref{fig:columbia}, where we show the
$(T-\mu_B)$-phase diagram of the (2+1)-flavor theory with $m_s=m^{\rm
  phys}$, in dependence of the light quark mass $m_{u,d}$. The shown
scenario assumes that $m_s^{tri}$ grows with $\mu_B$, implying that
the second order transition in the chiral limit turns into a first
order transition at some value of the chemical potential
$\mu_B=\mu_B^{tri}$.  A possible $Z(2)$-critical endpoint at physical
quark masses would then be smoothly connected with the tri-critical
point in the chiral limit by a line of second order transition. A
different scenario of a shrinking first order region is supported by
lattice simulations at imaginary chemical potential with
std. staggered fermions on coarse lattices \cite{deForcrand:2008vr}. In
this scenario, it is unlikely that a possible end-point at physical
quark masses can be smoothly connected to the critical surface of
$Z(2)$ end-points in the low mass limit of the three flavor theory.

Direct searches for the critical end-point on the lattice are so far
not very conclusive. First studies that reported a positive result
\cite{Fodor:2004nz, Fodor:2001pe} used a finite volume scaling analysis of
the Lee-Yang zeros together we a reweighting approach.  It is, however
not clear whether this method could produce fake signals \cite{Ejiri:2005ts}. 
Later on, a number of positive results have been
reported from calculations with the canonical partition function 
\cite{deForcrand:2006ec, Ejiri:2008xt, Li:2010qf, Li:2011ee}, 
which suffer from small volumes, large cut-off
effects and large quark masses. More recent calculations are based on 
estimating the radius of convergence of the Taylor expansion in $\mu_B$ 
of some suitable observables \cite{Allton:2003vx}. The problems here are twofold,
first, only a few expansion coefficients are available which makes the 
estimate of the radius of convergence difficult. Second, the first few
expansion coefficients are in agreement with calculations from the 
hadron resonance gas model for $T<Tc$. This models exhibits an infinite
convergence radius and thus no critical point. Nevertheless two groups report
estimates for the convergence radius \cite{Datta:2016ukp, DElia:2016jqh}.

\subsection{Universal scaling in (2+1)-flavor QCD}
\begin{figure}
\begin{center}
\hspace*{1.5mm}\includegraphics[height=0.325\textwidth, width=0.458\textwidth]{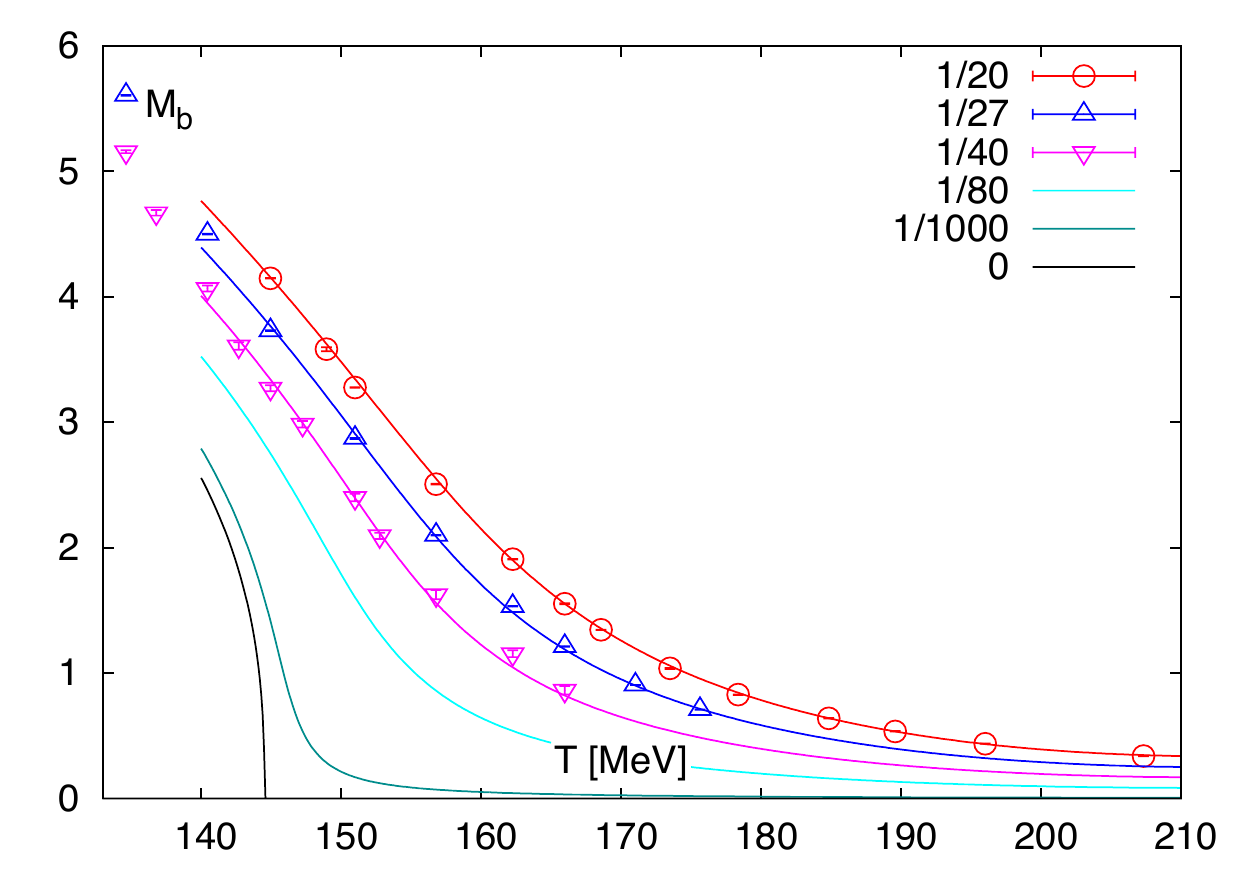}\hspace*{-0.5mm}
\includegraphics[height=0.325\textwidth, width=0.445\textwidth]{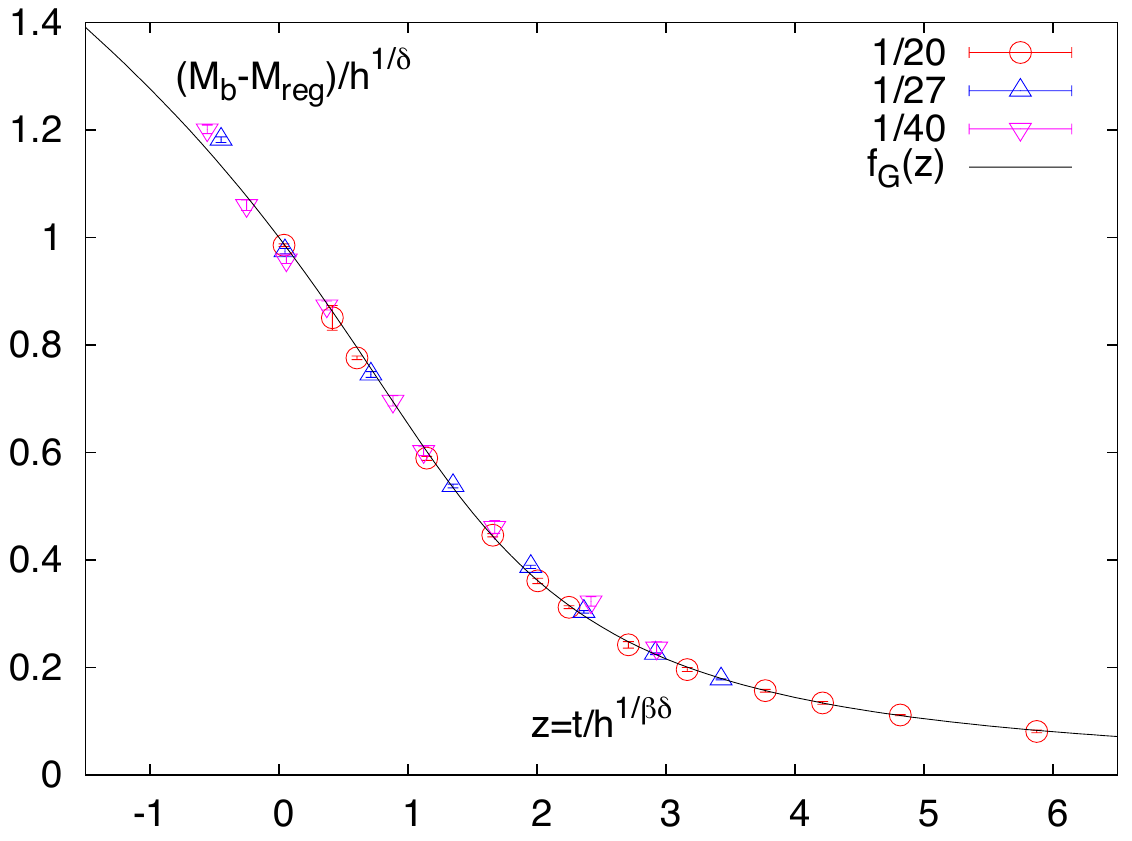}
\includegraphics[height=0.325\textwidth]{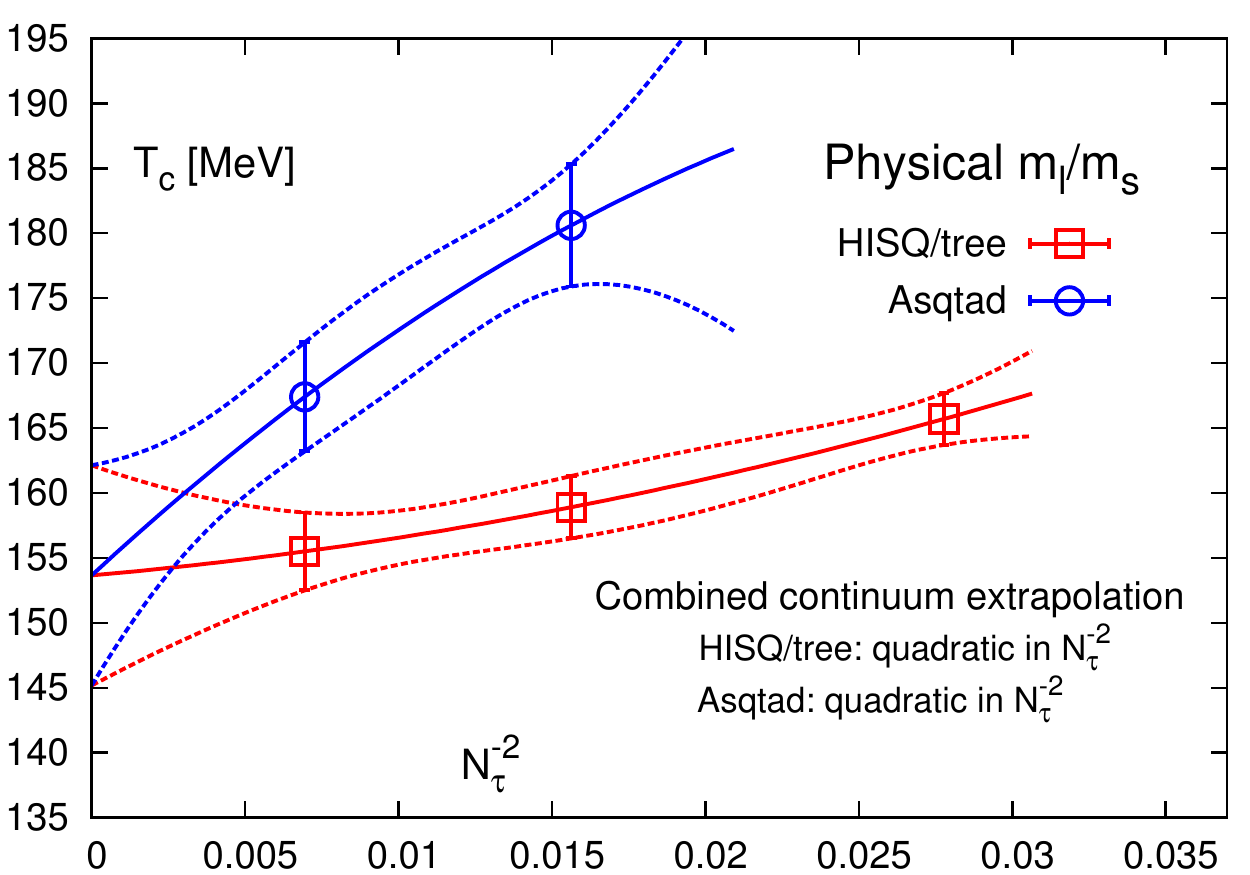}
\includegraphics[height=0.325\textwidth]{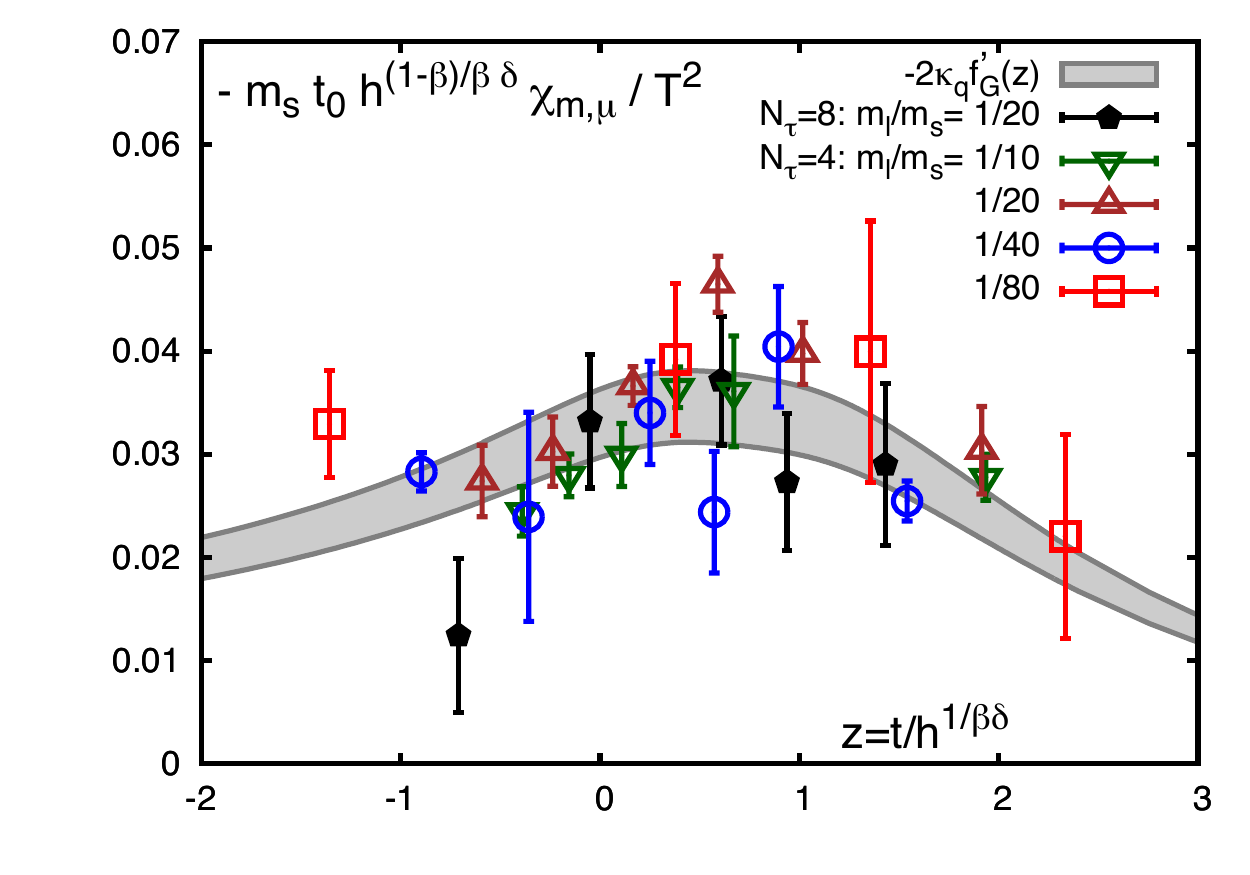}
\end{center}
\caption{Upper row: A scaling fit of the chiral condensate ($M_b$) to the magnetic 
equation of state using the $O(4)$ scaling function $f_G$, Eq.~(\ref{eq:fG}). Lower row: Continuum extrapolation of the 
transition temperature at physical quark masses performed jointly with two different lattice discretization (left) and a scaling 
fit the the mixed susceptibility $\chi_{m,\mu}$ (right), from Ref.~\cite{HotQCD_Tc, Kaczmarek:2011zz}.
\label{fig:scaling}}
\end{figure}

According the the scaling hypothesis, 
the free energy has a singular part that is responsible for the power
laws that thermodynamic response functions exhibits near the critical 
point. For two degenerate light flavors of mass $m_l\equiv m_u=m_d$ and 
one heavier strange quark mass $m_s=m_s^{phys}$, we thus make 
the Ansatz
\begin{equation}
\frac{p}{T^4}=\frac{1}{VT^3}\ln Z(T,V,m_l,\vec\mu)=-f_s(T,V,m_l,\vec\mu)
-f_r(T,V,m_l,\vec\mu),
\end{equation}
where the singular part $f_s$ will become a generalized homogeneous function of its
arguments once the correct scaling fields have been chosen as its natural variables. 
In general, all $O(N)$ and $Z(N)$ models exhibit two relevant scaling fields, 
the temperature-like ($t$) and external field-like scaling fields ($h$). Considering 
a critical point in the chiral limit ($m_l=0$) and at zero chemical potentials ($\vec\mu=0$) 
the leading order dependence of the scaling fields on $T,m_l,\vec\mu$ are determined by 
symmetry arguments, {\it i.e.} to leading order $h$ depends only on couplings that break 
chiral symmetry in the light quark sector, while $t$ depends on all other couplings. We find
\begin{equation}
t\equiv \frac{1}{t_0}\left(\frac{T-T_c}{T_c}+\kappa_l\hat\mu_l^2+\kappa_{ls}\hat\mu_l\hat\mu_s+\kappa_s\hat\mu_s^2\right)
\qquad \mbox{and} \qquad
h=\frac{1}{h_0}\frac{m_l}{m_s},
\end{equation}
where $T_c$ is the phase transition temperature and $t_0,h_0$ are non-universal scale 
parameters. Here we have adopted the $\vec\mu=(\mu_q,\mu_I,\mu_s)$ basis and assume $\mu_I=0$ as
a finite isospin chemical potential $\mu_I$ would break the flavor symmetry in such a way that a 
different symmetry breaking pattern would arise. With the above scaling fields we can 
exploit the property of homogeneity of the function $f_s$ to obtain
\begin{equation}
f_s(t,h)=h_0 h^{1+1/\delta}f_f(z)= h_0 h^{(2-\alpha)/\beta\delta}f_f(z) \qquad \mbox{with} \qquad z=t/h^{1/\beta\delta}.
\end{equation}
 The function $f_f$ is a universal scaling function that depend on a single scaling variable $z$. In this way we have 
singled out the leading order singular behavior, but have neglected sub-leading terms that are produced by 
irrelevant scaling variables and are known as corrections to scaling. From here we can easily obtain the magnetic
equation of state $M=h^{1/\delta}f_G(z)$ by taking a derivative with respect to $H=h_0h$,
where the scaling function $f_G$ is connected to $f_f$ by 
\begin{equation}
f_G(z)= - \left(1+\delta^{-1} \right)f_f(z) + z \left(\beta\delta\right)^{-1} f_f^{\prime}(z) \;.
\label{eq:fG}
\end{equation}
Note that this scaling form is not the famous Widom-Griffiths from, as the scaling variables
do not depend on the Magnetization. It is, however, well suited for a comparison with QCD where  
the calculated magnetization, which can, {\it e.g.}, be defined as $M_b=m_s\left<\bar\psi\psi\right>/T^4$ has
statistical and systematical errors. As an example of a scaling fit to the magnetization $M_b$ see the upper
part of Fig.~\ref{fig:scaling}.
The lattice data was obtained with the HISQ action on $N_\tau=8$ lattices and quark mass ratios 
$m_l/m_s=1/20,1/27,1/40$ \cite{Bazavov:2011nk}. The fit parameter are  $t_0$, $h_0$ and $T_c^0$ and 
three additional parameter for the regular part. The right panel shows the data collapse obtain when proper 
scaling variables are used. It can be seen that the scaling Ansatz works fine for quark masses as large as 
$m_l/m_s=1/20$. The relative contribution of the regular part has been found to be small. 

A more detailed scaling analysis was performed in Ref.~\cite{Bazavov:2011nk}, using a joint fit to the chiral condensate and 
chiral susceptibility. As a result we show in the lower left panel of Fig.~\ref{fig:scaling} the continuum extrapolation of the transition
temperature at the physical mass point, performed jointly with two different lattice discretization schemes, HISQ and asqtad. The obtained value
is $T_c=154\pm 9$ $\mbox{MeV}$. Other studies with improved staggered fermions, that use also other criteria for the definition 
of the crossover temperature, which are not directly related to universal critical behavior obtain similar values
 \cite{Aoki:2006br, Aoki:2009sc, Borsanyi:2010bp}. Also a study using chiral Domain Wall Fermions \cite{Bhattacharya:2014ara} yields 
 a compatible value. For the value for the phase transition temperature in the chiral limit, we obtain values of the oder of $T_c^0\approx 145$ 
 $\mbox{MeV}$.

Given the parameters $h_0,t_0$, we are able to determine the non
universal parameter $\kappa_l$ as well, which is the curvature of the
phase transition temperature in the chiral limit. The scaled
susceptibility $m_s t_0 h^{(1-\beta)/\beta\delta}\chi_{m,\mu}/T^2$ can
be fitted to the scaling function $f_G^\prime(z)$, with $\kappa_l$ being 
the only parameter left, which enters as a normalization factor.
We obtain $\kappa_q=0.059(5)$ \cite{Kaczmarek:2011zz}, which translates 
into a curvature of $\kappa_B=0.0066(5)$, associated 
with a change of the baryon chemical potential $\mu_B$. This result is 
not yet continuum extrapolated, but ultimately linked the critical behavior 
in the chiral limit. Other results based on imaginary chemical potential or
a Taylor expansion of chiral and deconfining order parameter are in the range 
$0.007<\kappa_B<0.02$ \cite{Bonati:2015bha, Bellwied:2015rza, Cea:2015cya}.

\input{./dof}

\input{./freeze}

\section{Summary and outlook}
We have reviewed selected results on the QCD phase structure and bulk
thermo\-dynamics, obtained by recent lattice calculations. We have
discussed results on derivatives of the logarithm of the QCD partition
function ($\ln Z$) with respect to
temperature and quark mass ($T,m$) and various chemical potentials
$\vec\mu$.  Results on correlation-functions have been not touched and
are beyond the scope of this review.

We have argued that the QCD equation of state (EoS), {\it i.e.}  the
pressure, energy and entropy density as a function of temperature,
obtained by the integral of the temperature derivative of $\ln Z$, 
has been reliably determined in the temperature
interval of $(130-400)$ MeV. The EoS is the basis for every hydro
dynamical model of the fireball created in heavy ion collisions. We
find that its low and high temperature limits seem to be in good
agreement with HRG model calculations and (re-summed) perturbative QCD
calculations, respectively. However, more refined calculations in both
of these temperature limits are needed.

We have shown that the behavior of the chiral condensate and (mixed)
chiral susceptibility exhibit universal critical behavior in the
chiral limit of ($2+1$)-flavor QCD, which is taken by sending the
quark mass of the two light flavors to zero while keeping the strange
quark mass constant. Matching QCD to the universal theory through the
determination of a set of non-universal parameter, yields valuable
informations on the QCD phase diagram. As an example we have discussed
the pseudo-critical line as function of $m$ and $\mu_B$. Although we
find a scaling behavior that is consistent with an $O(4)$ symmetric 
critical point in
the chiral limit, the possibility of a first order transition in the
chiral limit and correspondingly a critical $Z(2)$ end-point at very small
but finite quark masses is not excluded yet. This issue is closely
related with the possibility of a sufficiently restored axial anomaly
($U_A(1)$-symmetry) close to $T_c$, which is currently subject to many
lattice QCD calculations with chiral fermions. Most studies 
do, however, support the opposite scenario that $U_A(1)$ is strongly 
broken around $T_c$.

Finally we have discussed derivatives of $\ln Z$ with respect to
various chemical potentials (generalized susceptibilities).  We have
introduced combinations of these generalized susceptibilities which
allow for the disentanglement of the partial pressures of different
hadronic sectors and thus lead to a better understanding of the
relevant microscopic degrees of freedom. We presented evidence for
experimentally not yet measured strange and charmed hadrons and showed
that also flavored hadrons start to melt at the chiral transition
temperature. Vise versa, we used similar combinations of these
generalized susceptibility to investigate the approach to the free gas
behavior. We find that the temperature range of $1<T/T_c<2$ can be
characterized by a strongly interacting plasma of quarks, gluons and
some remnants of hadron like states, while above $T/T_c\gtrsim 2$ the
quasi-particle spectrum resembles that of quarks and gluons alone.

The same generalized susceptibilities also encode the fluctuations of
conserved charges, which are measured at RHIC and LHC. We discussed
certain ratios of cumulants of conserved charges and showed that from a
fit to the corresponding experimentally measured data, the freeze-out
temperature, as well as the freeze-out curvature can be obtained. We
find that the so determined freeze-out line is in good agreement with
current determinations of the pseudo-critical line.

We emphasize that several of the $4^{\rm th}$ and $6^{\rm th}$ order
susceptibilities that are used in the above summarized studies are not
yet continuum extrapolated. Although we do not expect large cut-off
effects for these observables, continuum extrapolations are necessary
to consolidate our findings and are currently work in progress. This
will also allow to extent the range of the QCD equation of state to
$\mu_B/T\lesssim 2$, which corresponds to the full range of the energy
scan program at RHIC \cite{eosmu}. Even though some results on $8^{\rm
  th}$ order susceptibilities are already available and will be refined in future, it is
evident that there are many interesting aspects on the QCD phase
diagram that will never be in reach by a Taylor expansion method.
Therefore a reformulation of QCD in terms of other variables that
allows for direct simulations at nonzero chemical potential will be
needed.

\section*{Acknowledgment}
We thank all members of the Bielefeld-BNL-CCNU collaboration for
valuable discussions and comments. S.S. acknowledges support
by the U.S. Department of Energy under Grant No. DE-SC0012704.

\end{document}

%% file: dof
\section{The degrees of freedom in QCD near deconfinement
\label{sec:dof}}
An important question in QCD is to understand what happens to the
microscopic degrees of freedom at deconfinement. For the light quark
sector in QCD, lattice studies point to the fact that colored degrees
of freedom start to appear already when the chiral symmetry is
approximately restored \cite{HotQCD_Tc}. Since the strange quark mass
is $\sim \Lambda_{QCD}$, it is not a priori obvious whether the same
phenomena would be true even for the strangeness sector. Also of
particular interest are the bound states of charm quark. It is now
known from the first lattice studies
\cite{Asakawa:2003re,Datta:2003ww} that $c\bar c$ bound states like
the $J/\psi$ may survive in the hot QCD medium upto $1.4~T_c$. The
interesting problems to understand non-perturbatively are what is the
fate of hadrons consisting of charm and light quarks near $T_c$, and
what are the dominant degrees of freedom carrying charm in the
deconfined medium.

\subsection{Fate of strange hadrons}
\label{melts}
\begin{figure}
\begin{center}
\includegraphics*[scale=0.45]{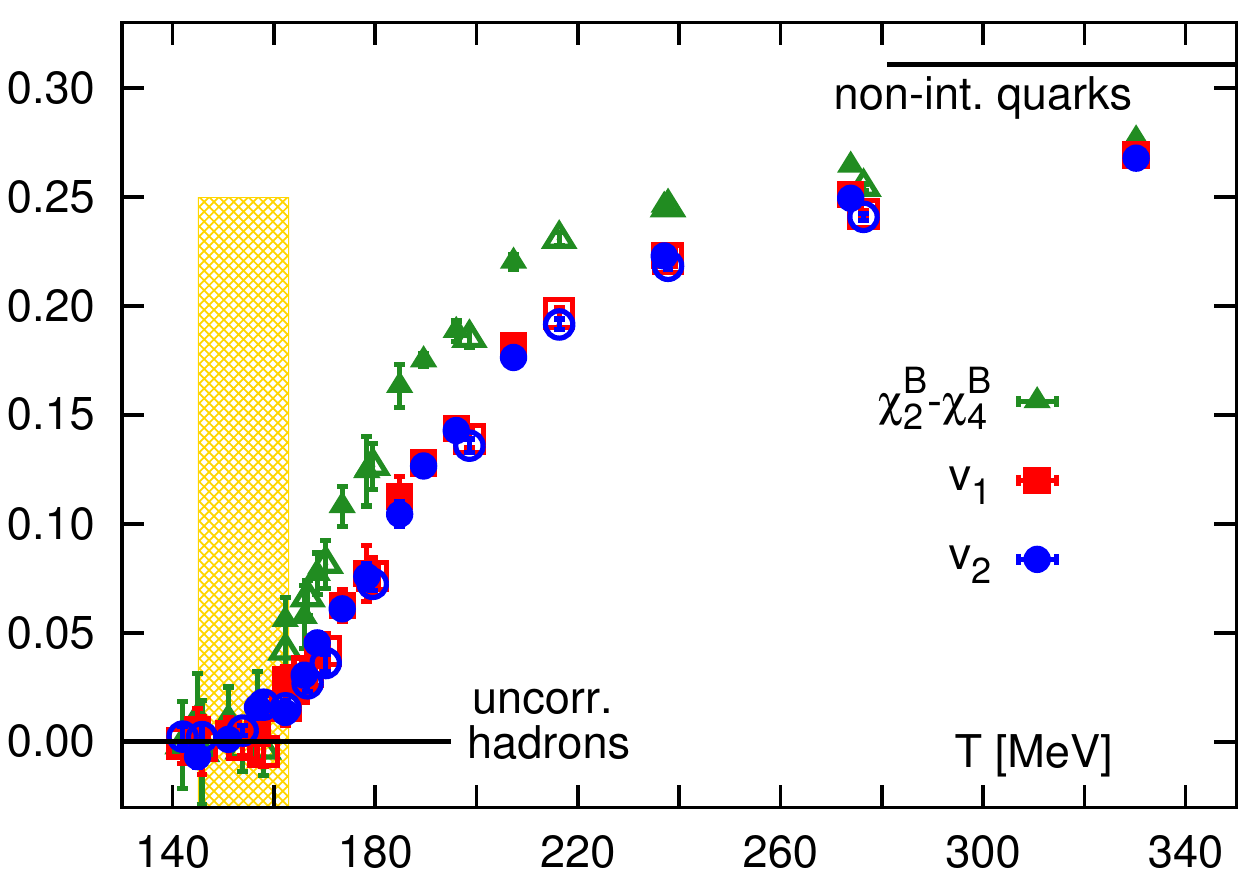}
\includegraphics*[scale=0.45]{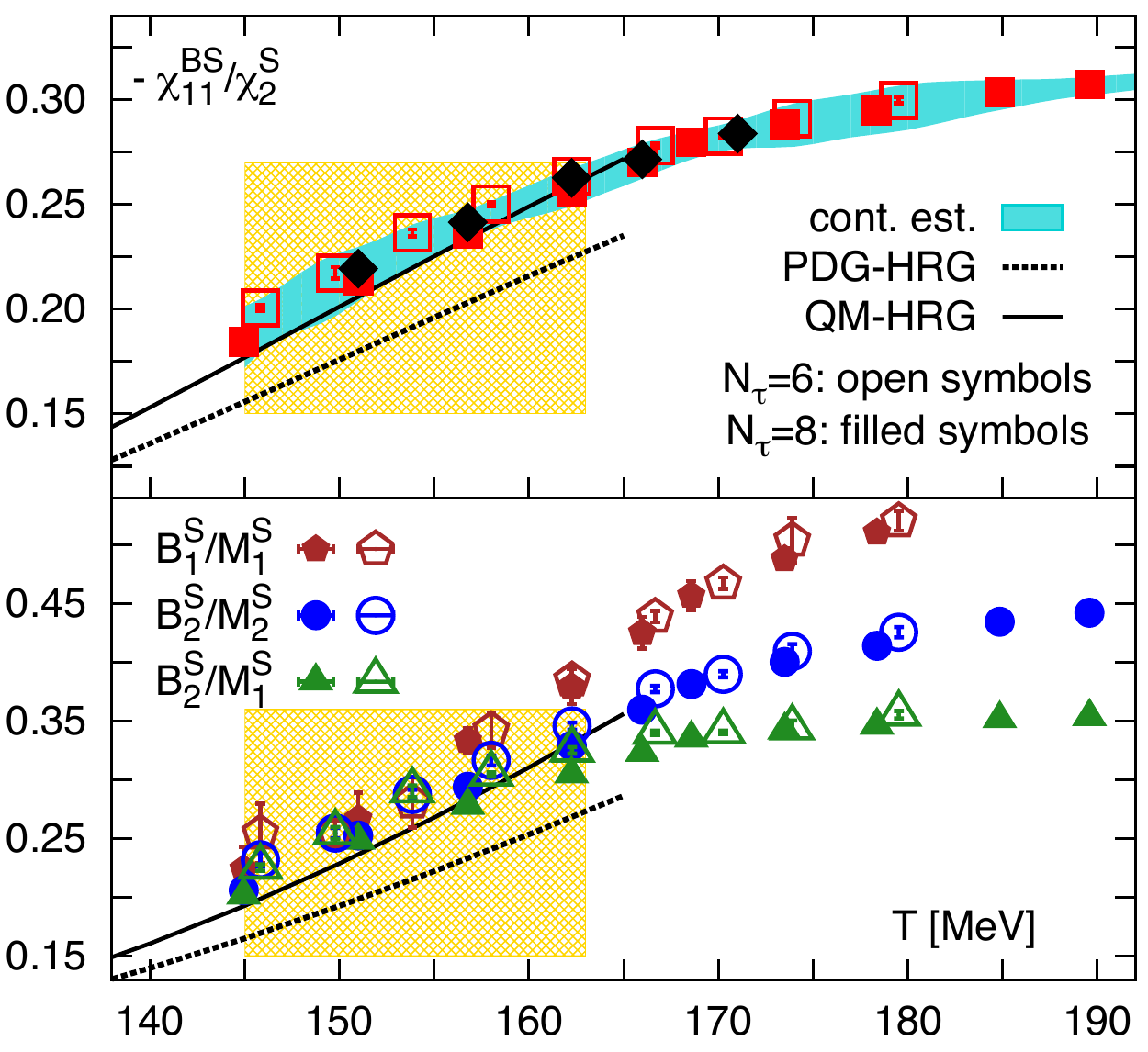}
\caption{The left panel shows the constraints for the HRG inspired
  model from Ref. ~\cite{Bazavov:2013dta} for strange hadrons
  calculated on a $24^3\times6$ (open symbols) and $32^3\times 8$
  (closed symbols) lattice. The right panel shows the comparison of
  the ratio of partial pressures for strange baryons to the total
  strange pressure to the HRG model taken from
  Ref.~\cite{Bazavov:2014xya}.}
\label{fig:meltings}
\end{center}
\end{figure}

Below $T_c$, thermodynamic quantities like the pressure in QCD can be
fairly well described as consisting of non-interacting hadrons and
resonances. The total pressure of a thermal ensemble of a
non-interacting gas of mesons, baryons and resonances containing
strangeness quantum number $S$ at finite chemical potentials
$\hat\mu_X$ $=$ $\mu_X/T$, $X\equiv B,S$ is given as,
\begin{eqnarray}
\nonumber
 P(\hat\mu_S,\hat\mu_B)&=&P_M\cosh (\hat\mu_S)+P_{B,S=1}\cosh(\hat\mu_B+\hat \mu_S)+ \\
&& P_{B,S=2}\cosh(\hat\mu_B+2\hat \mu_S) +P_{B,S=3}\cosh(\hat\mu_B+3\hat \mu_S)~.
\end{eqnarray}
where $P_M$ and $P_{B,S=n}$ denote the partial pressure of
open-strange mesons and re\-spectively baryons containing strangeness
quantum number $n$ at vanishing chemical potentials. The derivatives
of $P(\hat\mu_S,\hat\mu_B)$ with respect to $\mu_X$, $\chi_{kl}^{BS}$ $=$
$\left.  \frac{\partial^{(k+l)} [P(\hat\mu_B,\hat\mu_S)/T^4]} {\partial
  \hat\mu_B^k \partial \hat\mu_S^l} \right|_{\vec{\mu}=0}$, define
different generalized susceptibilities which encode the fluctuations
of $B,S$ and their correlations at vanishingly small chemical
potentials. In a non-interacting gas of strange hadrons, these
quantities can be written as,
\begin{eqnarray}
\chi_n^S(\hat\mu=0)&=&P_M+1^n P_{B,S=1}+2^n P_{B,S=2}+ 3^n P_{B,S=3}~,~\\
\chi_{kl}^{BS}(\hat\mu=0)&=&1^l P_{B,S=1}+2^l P_{B,S=2}+ 3^l P_{B,S=3}~.
\end{eqnarray}
Inverting above relations would then give us the partial pressures in
the meson and baryon sectors, $P_M, P_B$ at $\mu_X=0$
\cite{Bazavov:2013dta}. The crucial input for this calculation are the
susceptibilities $\chi_{kl}^{BS}$ measured on the lattice. If one
measures susceptibilities upto fourth order, then in the $B$-$S$
sector one has $6$ independent measurements $\chi_2^S,\chi_4^S,
\chi_{11}^{BS}, \chi_{22}^{BS},\chi_{31}^{BS},\chi_{13}^{BS}$. If
indeed the HRG model describes the data, then one could extract the
$4$ independent parameters of the model which are $P_M$ and three
$P_B$'s in terms of these susceptibilities, along with $2$ independent
constraint equations. The lattice QCD data should satisfy the
constraints imposed on the model, in order to validate the model. In
terms of the susceptibilities, the $2$ independent constraints have
been derived in Ref. \cite{Bazavov:2013dta},
\begin{equation}
 v_1=\chi_{31}^{BS}-\chi_{11}^{BS}~,~
 v_2=\frac{\chi_2^S-\chi_4^S}{3}-2 \chi_{31}^{BS}-4 \chi_{22}^{BS}- 2 \chi_{13}^{BS}~.
\end{equation}
The values of $v_1$ and $v_2$ based on measurements on $N_\tau=6,8$
lattices with obtained with the HISQ action from
Ref.~\cite{Bazavov:2013dta} are summarized in
Fig. \ref{fig:meltings}. It can be concluded~\cite{Bazavov:2013dta}
that the HRG description for strange hadrons starts to break down at
the chiral crossover transition region. In fact, the ratio of the
partial pressure of strange baryons to the total HRG pressure given by
the quantity $-\chi_{11}^{BS}/\chi_2^S$ shown in the right panel of
Fig.~\ref{fig:meltings} from~\cite{Bazavov:2014xya} start to show
deviation from the HRG description already at $T_c$, implying
deconfinement of strangeness degrees of freedom. Similar consequences
for strange hadrons have been observed independently from a different
set of observables in Ref.~\cite{Bellwied:2013cta} motivated from the
above outlined analysis.

\subsection{Melting of open charm hadrons}
\label{meltc}
\begin{figure}
\begin{center}
\includegraphics*[scale=0.45]{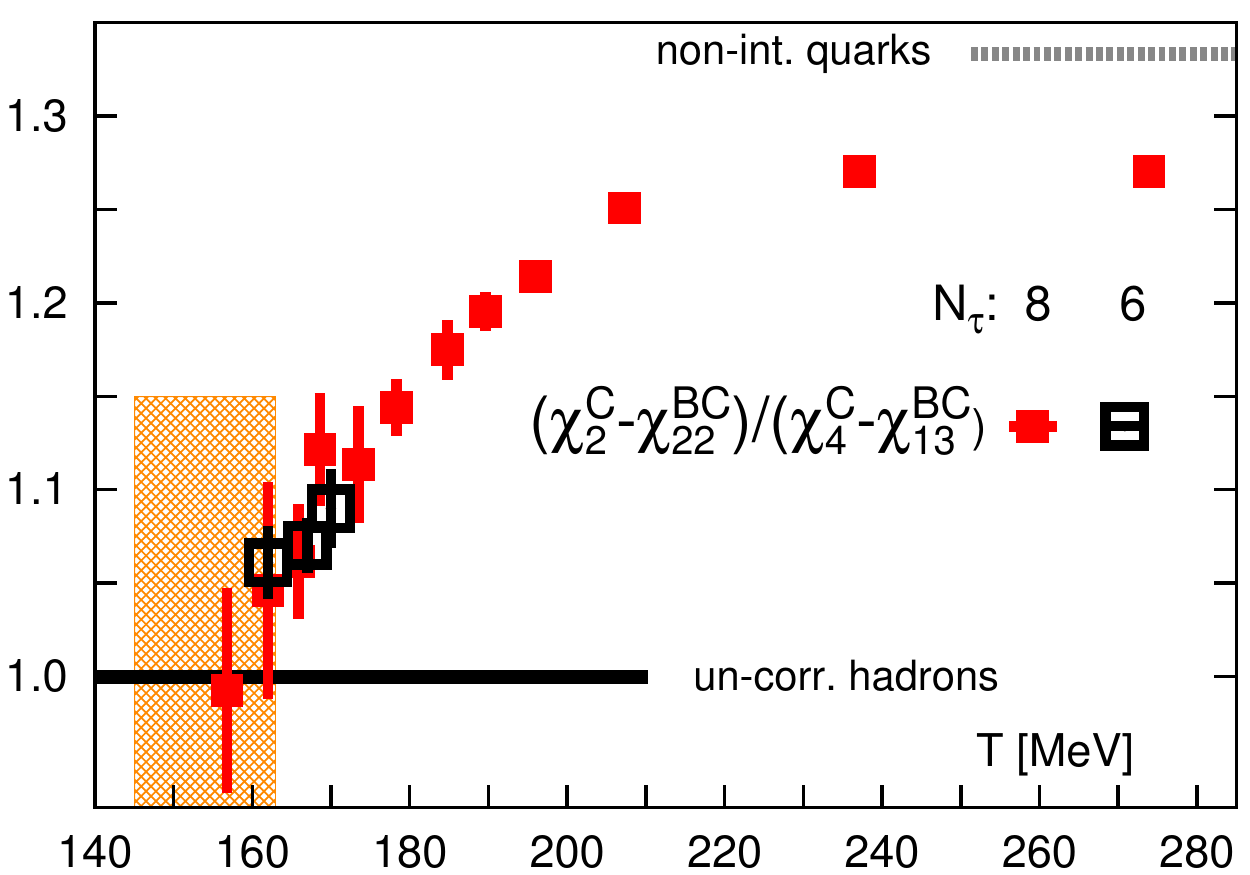}
\includegraphics*[scale=0.45]{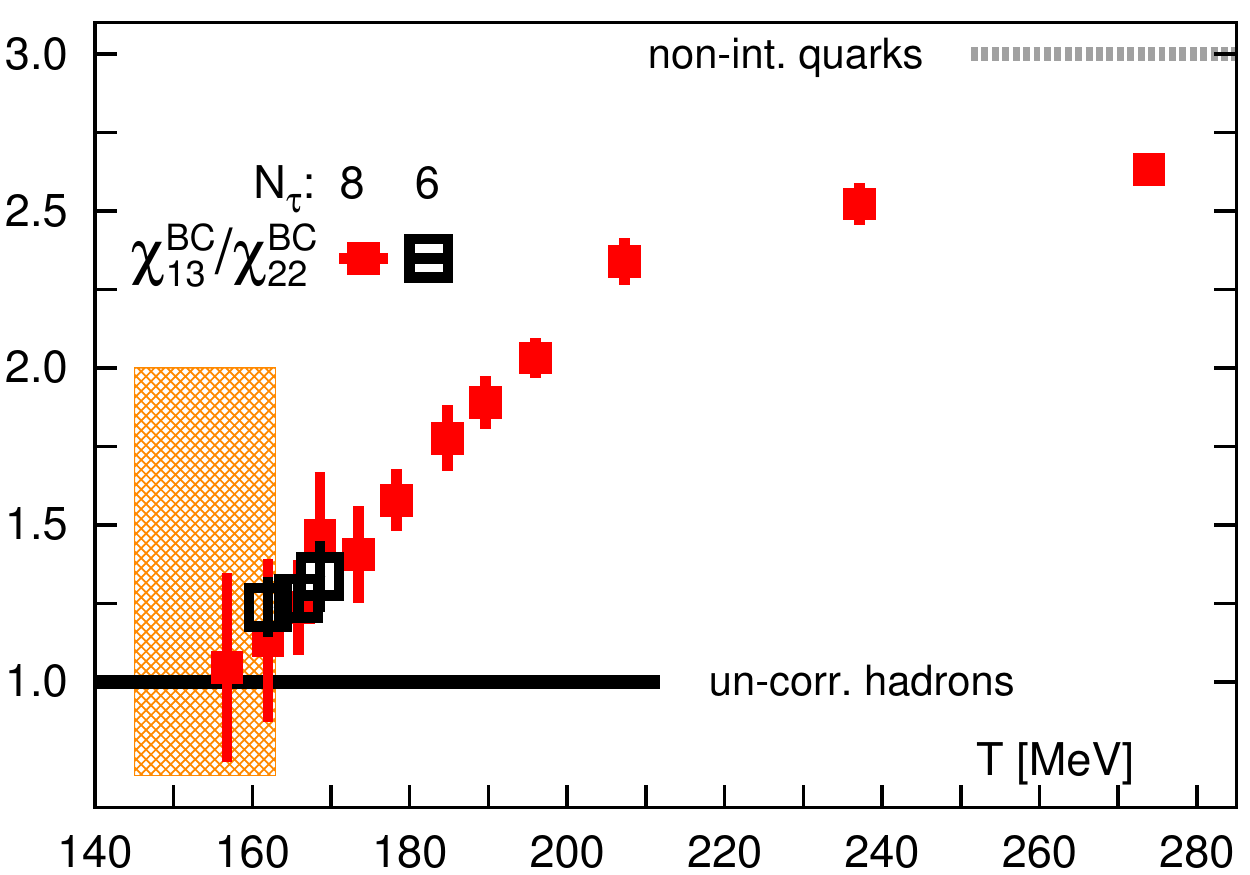}
\end{center}
\caption{The ratios of the partial pressure of the open charm
  mesons(left panel) and baryons(right panel) shown as function of
  temperature for two different lattice sizes with $N_\tau=6,8$ from
  Ref.~\cite{Bazavov:2014yba}.  The yellow band represents the
  crossover region in QCD. }
\label{fig:melting1}
\end{figure}

For the charm sector, the ground states of the baryons carrying charm
quantum numbers $C=2,3$ respectively, are heavier than the $C=1$
baryons by about $1$ GeV, and hence their partial pressure
contributions are suppressed near $T_c$. As a consequence the total
pressure of the charm degrees of freedom in the hadronic medium is
simply due to the sum of the partial pressures of charmed mesons $P_M$
and $C=1$ baryons $P_{B,C=1}$ respectively~\cite{Bazavov:2014yba}.
The $P_M$ can be expressed as two equivalent combinations of the
susceptibilities,
$P_M=\chi_2^C-\chi_{22}^{BC}=\chi_4^C-\chi_{13}^{BC}$. For the heavy
quarks, additional complication arise due to strong mass dependent
cut-off effects in the susceptibilities.  Instead, it has been
proposed in Ref.~\cite{Bazavov:2014yba} that the ratio of the two
independent measures of $P_M$ would be a good observable with
significantly less contamination from the cut-off effects of order
$\mathcal{O}(ma)$.  The ratio should be unity in the hadron phase and
any deviation from unity would be a signal for the melting of these
charmed mesons. In the baryon sector, the relative contribution of the
partial pressures of $C=2,3$ baryons given by $\chi_4^C-\chi_2^C$ has
been indeed shown to be vanishingly small compared to $P_{B,C=1}$
~\cite{Bazavov:2014yba} near $T_c$. This considerably simplifies the
partial pressure due to the baryons carrying a unit charm to $ P_{B=1}
\simeq \chi_{mn}^{BC}~,m,n>0$ and $m+n=$ even. Hence, ratios like
$\chi_{13}^{BC}/\chi_{22}^{BC}$ would be unity in a phase consisting
of charmed baryons. The partial pressures for the charmed meson and
baryon sectors from Ref.~\cite{Bazavov:2014yba} are summarized in
Fig.~\ref{fig:melting1}.  It is evident that both open charm meson and
baryon partial pressure ratios depart from unity already in the
crossover region signaling the deconfinement of charm at $T_c$. The
conclusions based on these ratios are quite robust since these are
independent of the details of the hadron spectrum and have a very mild
lattice spacing dependence seen from the $N_\tau$ independence of the
results in Fig.~\ref{fig:melting1}.

\subsection{Evidence of additional charm hadrons from QCD thermodynamics}
\label{charmhrg}
\begin{figure}
\begin{center}
\includegraphics*[scale=0.45]{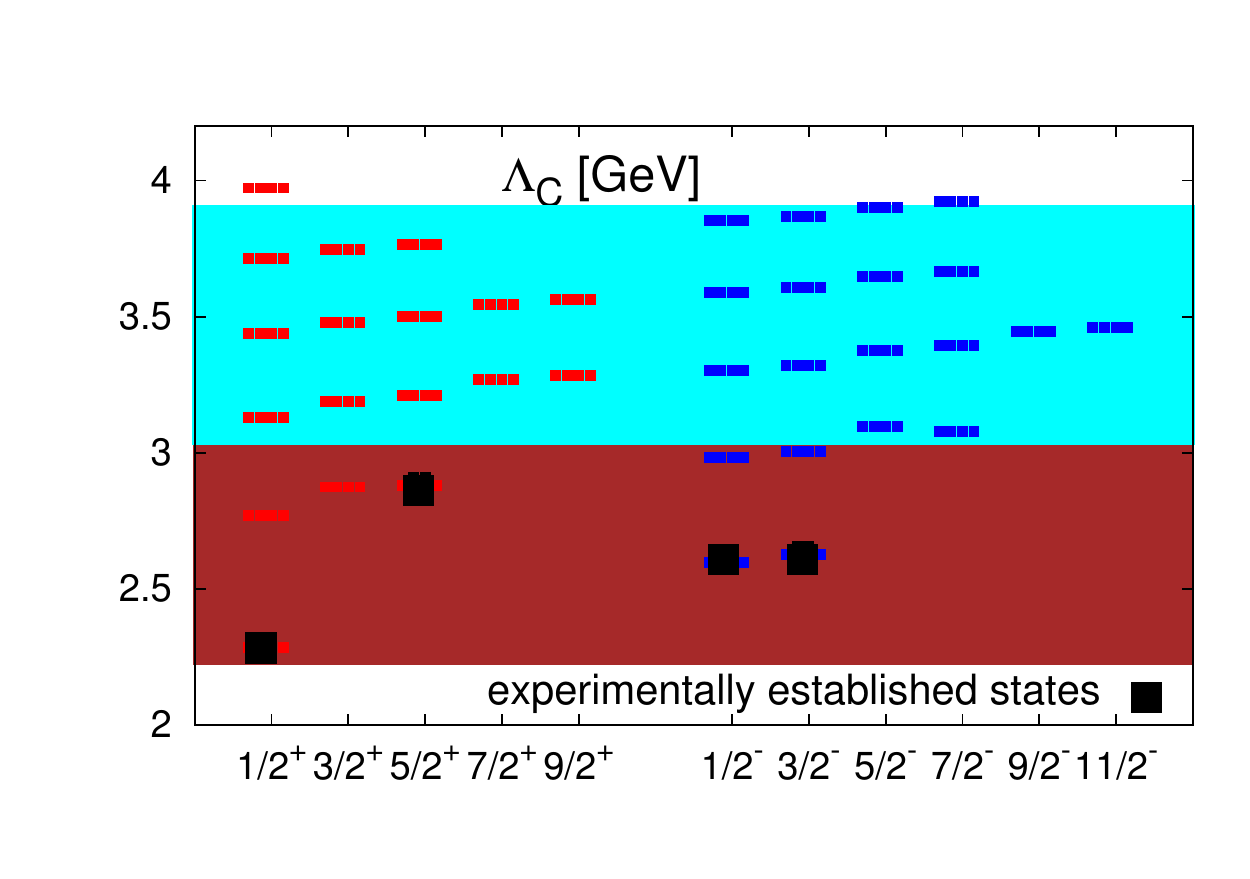}
\includegraphics*[scale=0.35]{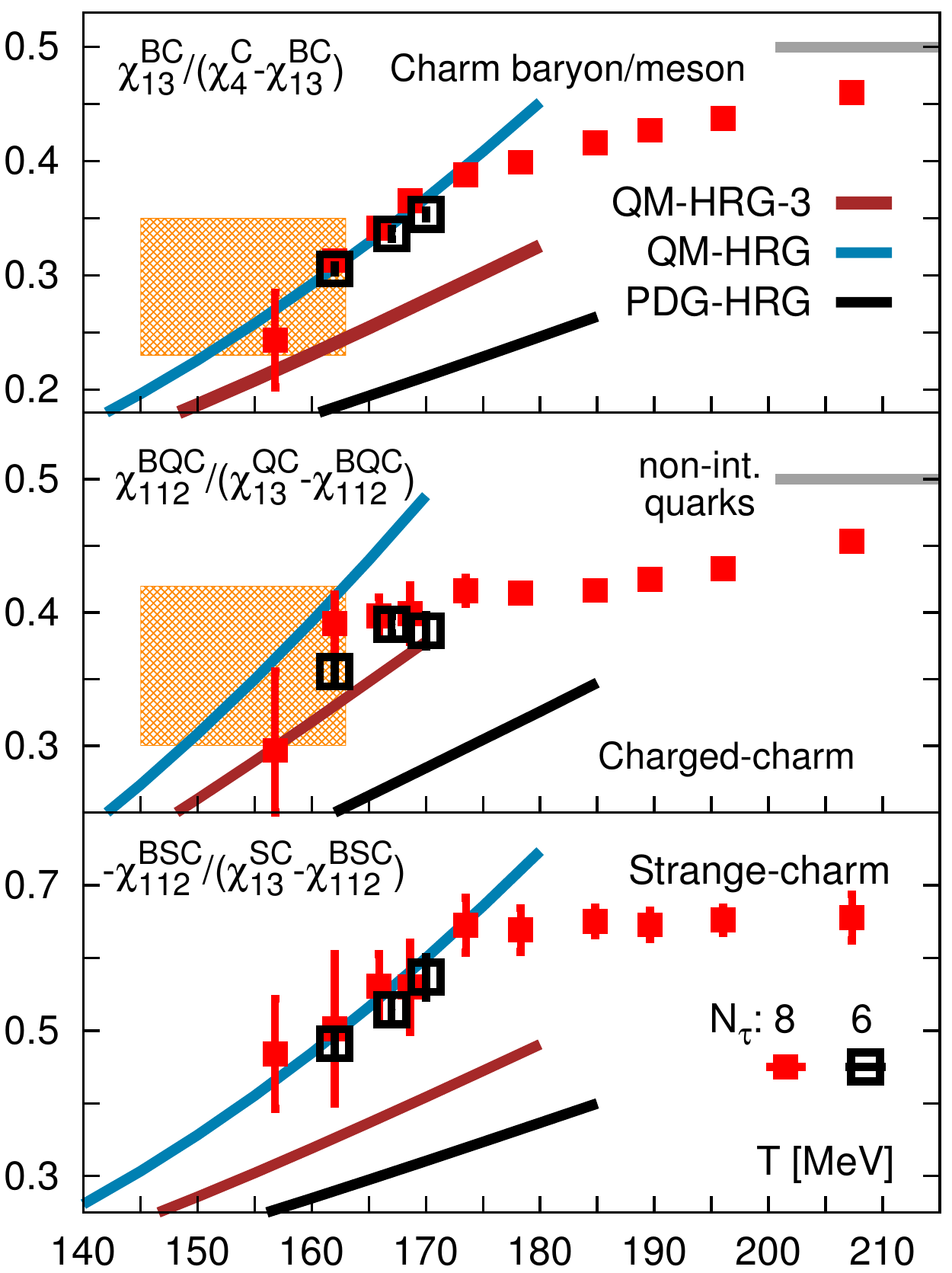}
\end{center}
\caption{The spectrum of $\Lambda_c$ from Quark Model calculations in
  the left panel has many more states that are not yet detected in the
  experiments. The brown band represents open charm hadrons with
  masses less than $3$ GeV and the blue band represents all such known
  states. In the right panel, our lattice QCD data of the ratio of
  partial pressures of the open charm mesons and baryons supports the
  existence of the additional charm baryon states in different
  channels which are predicted from the Quark Model and hadron
  spectrum studies on the lattice.}
\label{fig:charmhrg}
\end{figure}

Since these newly proposed combinations of generalized susceptibility,
proposed for the first time in
Refs.~\cite{Bazavov:2013dta,Bazavov:2014yba} and discussed in
Sec.~\ref{melts} and \ref{meltc}, allow for the disentaglement of the
partial pressures of meson and baryon sectors with different quantum
numbers, we can further understand the microscopic composition of
individual sectors at the chiral crossover transition.  Calculations
in the Quark Model(QM)~\cite{Capstick:1986bm,Ebert:2011kk} as well
hadron spectrum studies on the
lattice~\cite{Padmanath:2013bla,Padmanath:2015bra} predict many more
charmed baryons than experimentally detected and tabulated in the
Particle Data Table~\cite{Beringer:1900zz}, whereas for the charmed
meson sector there is quite a good
agreement~\cite{Ebert:2009ua,Moir:2013ub}.  For instance, excited
states for the $D, D_s$ mesons which are upto $700$ MeV above the
ground state are fairly well measured experimentally. On the other
hand, the spectrum of $\Lambda_c$ baryon is not well
measured~\cite{Bazavov:2014yba} as shown in the left panel of
Fig.~\ref{fig:charmhrg}.
Only three states above the ground state have been measured in the
experiments and assigned to a well defined spin and parity. However,
the ground state spin is not yet measured with certainty. There are
many more bound states in different spin-parity channels predicted
from the QM. Similar results are obtained from charm baryon
spectroscopy on the lattice~\cite{Padmanath:2015bra}. The presence of
these additional baryon states and resonances have been shown to have
a measurable impact on the pressure of open-charm hadrons near $T_c$
~\cite {Bazavov:2014yba}. The baryon states which have masses $1$ GeV
or more than the ground state lying above the blue band in
Fig.~\ref{fig:charmhrg} may be unstable or would have negligible
contribution to the partial pressures. For a systematic understanding
of the relative contributions of these states by comparing to the
lattice data, three different sets have been
proposed~\cite{Bazavov:2014yba}. These consist of a hadron resonance
gas(HRG) with all predicted states in the Quark Model (QM-HRG)
~\cite{Capstick:1986bm,Ebert:2011kk} denoted by brown lines, a
QM-HRG-3 consisting of Quark Model states upto $3$ GeV mass and a
PDG-HRG model consisting only of the experimentally known
states. Using appropriate combinations of susceptibilities, the
partial pressures of these channels with specific quantum numbers have
been in derived in Ref.~\cite{Bazavov:2014yba}. For the charged
sector, charmed baryon and meson partial pressures are given as
$P_B\simeq\chi_{112}^{BQC}$ and
$P_M\simeq\chi_{13}^{QC}-\chi_{112}^{BQC}$ respectively whereas for
the charm baryons additionally carrying strangeness,
$P_B\simeq\chi_{112}^{BSC}$ and
$P_M\simeq\chi_{13}^{SC}-\chi_{112}^{BSC}$.  The comparative results
between lattice data and predictions from different HRG models are
summarized from Ref.~\cite{Bazavov:2014yba} in the right panel of
Fig.~\ref{fig:charmhrg}. In all these sectors a significant
contribution of the additional baryon states over PDG-HRG at the
crossover transition has been found.  This is a remarkable example
where finite temperature lattice QCD data provides an independent
motivation for further experimental and theoretical studies on charm
hadron spectroscopy~\cite{Padmanath:2015bra}.

\subsection{Nature of quasi-particles beyond \texorpdfstring{$T_c$}{Tc}}
\label{charmquasi}
Evidence from the results for screening masses in the charm
sector~\cite{Bazavov:2014cta} show that the baryon and meson-like
excitations do not immediately disappear at the chiral crossover
transition.  The screening masses only reaches the free fermion limit
at $T>250$ MeV. Moreover comparison of the lattice data for
off-diagonal susceptibilities like $\chi_{22}^{uc}$ normalized by
$\chi_2^c$, to perturbative results within Hard thermal Loop
resummation, show that these are in fairly good agreement only beyond
$1.2~T_c$~\cite{Mukherjee:2015mxc}.  This further hints to the fact
that the dominant degrees of freedom carrying charm quantum number
changes at around $1.2~T_c$ where quark degrees of freedom starts
becoming dominant in the plasma. Owing to the large charm quark mass,
the charm-carrying degrees of freedom could be well described in terms
of quasiparticles. It allows one to further decompose the total
pressure due to charm degrees of freedom into contributions from
baryons, mesons and quark-like excitations carrying unit charm quantum
number, following Ref.~\cite{Mukherjee:2015mxc},
\begin{eqnarray}
&
\displaystyle
P_{C}(T,\mu_B,\mu_C)=P_Q(T) \cosh\left(\frac{\mu_C+\mu_B/3}{T}\right)+\nonumber\\
&
\displaystyle
P_{B,C=1}(T) \cosh\left(\frac{\mu_C+\mu_B}{T}\right)+
P_M(T) \cosh\left(\frac{\mu_C}{T}\right).
\label{eqn:pc}
\end{eqnarray}
 \begin{figure}
\begin{center}
\includegraphics[width=5cm, height=5cm]{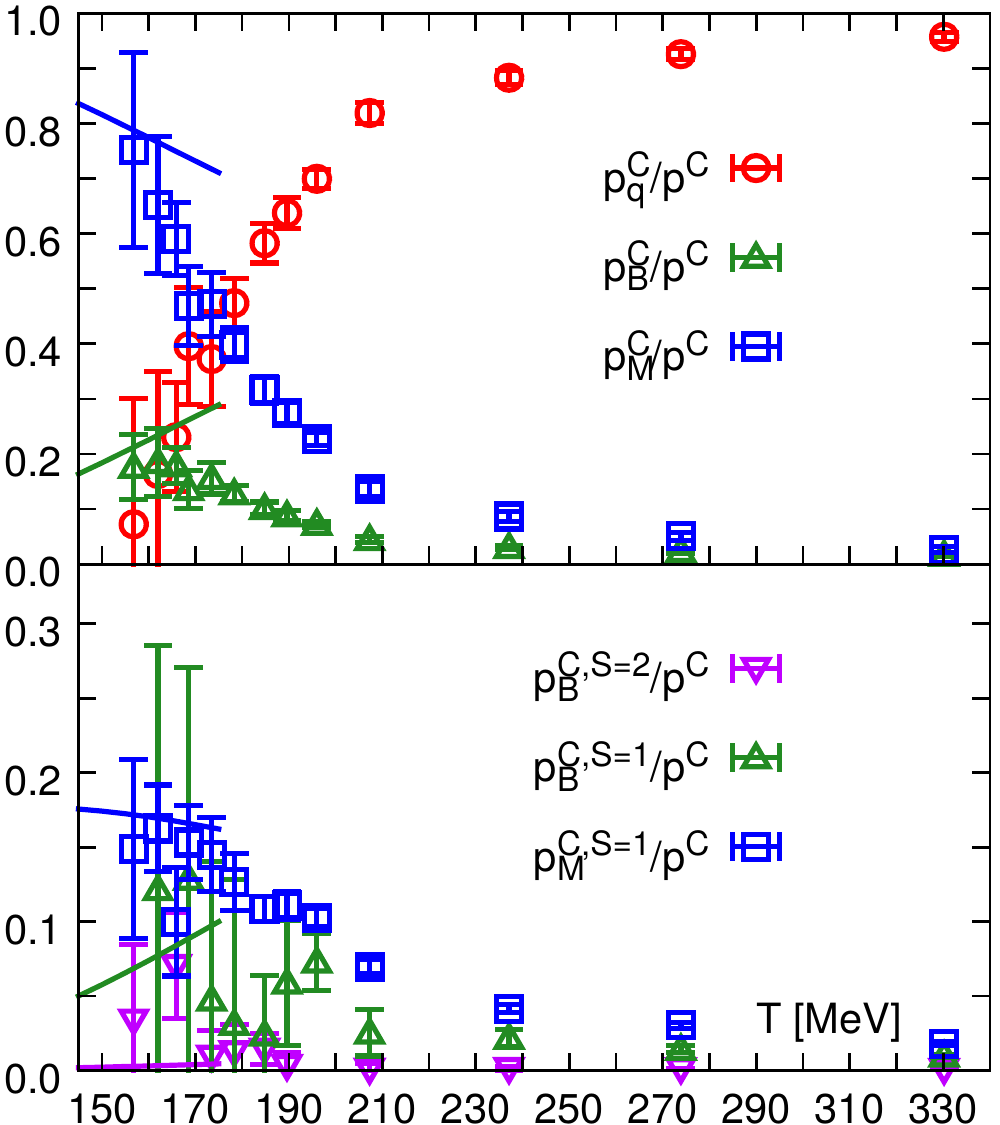}
\includegraphics[width=5cm, height=5cm]{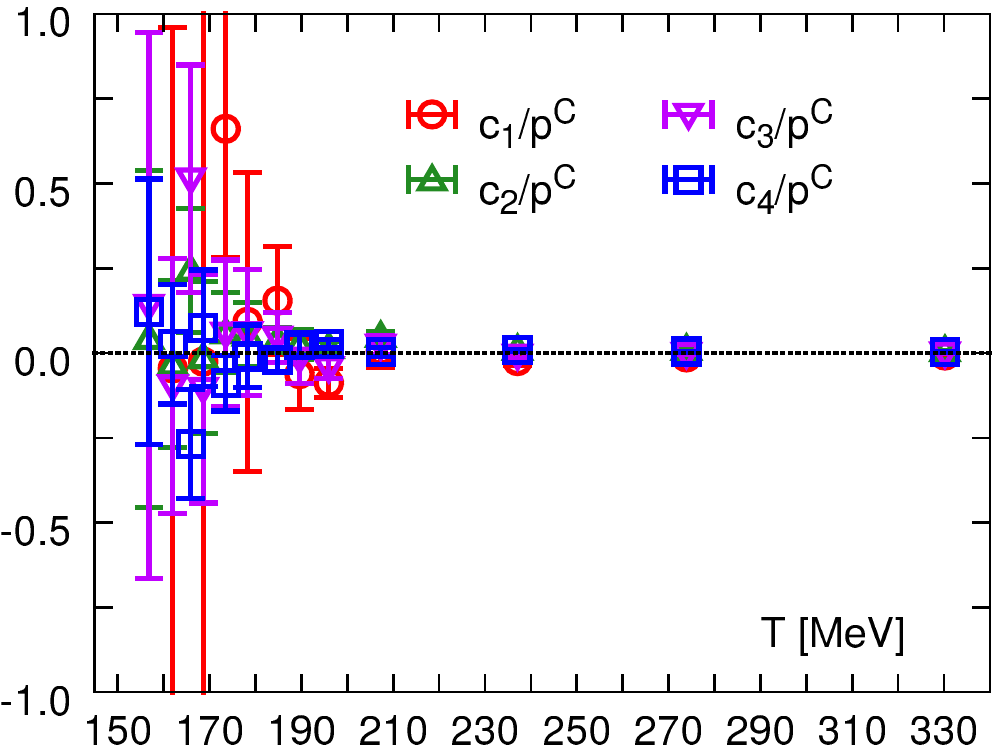}
\end{center}
\caption{The partial pressures of open charm mesons, charmed baryons and charm quarks as function 
of temperature ranging from $T_c$ to $2 T_c$ in the $BC$ and $SC$ sectors respectively (left panel). 
The solid lines show the corresponding partial pressures obtained from Quark Model HRG. In the right 
panel the lattice data is shown to satisfy the constraints imposed by different quasi-particle models.
Both the figures are from Ref.~\cite{Mukherjee:2015mxc}.}
\label{fig:charmpp}
\end{figure}

Again using the lattice data for charm correlations and fluctuations upto fourth order, the partial pressures $P_M$, $P_{B,C=1}$ and $P_Q$ can be solved, as outlined in Ref. ~\cite{Mukherjee:2015mxc} in terms of six observables $\chi_2^C$, $\chi_{11}^{BC}$, 
$\chi_{13}^{BC}$ $\chi_{22}^{BC}, \chi_{31}^{BC}, \chi_{4}^C$. The trivial constraints in the 
charm sector $\chi_4^C=\chi_2^C~,~ \chi_{11}^{BC}=\chi_{13}^{BC}~$, leaves only four independent 
observables. However, since three independent partial pressures need to be calculated within this 
model there would only be a unique non-trivial constraint for this model $c_1\equiv\chi_{13}^{BC}-4 
\chi_{22}^{BC}+3 \chi_{31}^{BC}=0$. The contribution of the three different partial pressures 
within this model has been calculated~\cite{Mukherjee:2015mxc}and summarized in Fig. \ref{fig:charmpp}. 
It is was concluded that the charm meson and baryon-like excitations contribute significantly to 
the total pressure till $T\sim 200$ MeV with the quark-like excitations dominating only for 
temperatures beyond. Moreover a specific sub-sector of this model was looked into detail~\cite{Mukherjee:2015mxc} consisting of quasi-particles which carry both charm and strangeness quantum 
numbers, since these two quantum numbers are completely decorrelated when quark-like excitations 
play a dominant role.   Indeed it was found that the meson and baryon-like excitations within this 
sector may survive well beyond $T_c$, with the meson-like contributions dominating~\cite{Mukherjee:2015mxc}. 
The quark-like excitations accounts for more than $50\%$ of the total pressure only around $200$ MeV. These studies also looked for 
evidences for the presence of di-quarks in the deconfined phase of QCD. Atleast for the charm sector, the presence of di-quark like excitations has not been found above $T_c$~\cite{Mukherjee:2015mxc}. 

To check whether indeed a model can ``mimic'' QCD, the  constraints that define the model should be exactly 
satisfied by the data from lattice QCD calculations. It was infact found in Ref.~\cite{Mukherjee:2015mxc} 
that the best available lattice data satisfies constraints for all the above mentioned models, which is shown 
in the right panel of Fig. \ref{fig:charmpp}. So in summary, there is a very strong evidence atleast for the 
charm sector, that the quark-like degrees of freedom start appearing already near $T_c$. However, these remain 
strongly correlated as color neutral baryon and meson-like excitations till about $1.2~T_c$. For the first 
time it was shown~\cite{Mukherjee:2015mxc} that colored quark-like excitations only start dominating 
thermodynamically beyond this temperature. This has important consequences for the phenomenological modeling 
of heavy-quark diffusion and quenching which needs the deconfinement temperature of the heavy-quarks and the 
nature and abundance of resonances that carry charm quantum number in QGP as crucial inputs~\cite{Prino:2016cni}.

%% file: freeze
\section{Freeze-out and connection to heavy ion experiments\label{sec:freeze}}
Many new insights on the nature of the quasi-particles in the hot QCD
medium and their interactions have been recently obtained from the
extensive lattice data on fluctuation measurements especially in the
charm sector~\cite{Mukherjee:2015mxc}.  As motivated in the
introductory sections, one of the outstanding issues is to understand
when does the HRG picture breaks down for QCD with light quarks. This
is not only important for our fundamental understanding of QCD, but it
also allows us to relate to the wealth of experimental data coming
from the heavy-ion experiments at Relativistic Heavy Ion Collider
(RHIC), BNL, and at CERN. In the experiments, the fireball created due
to the collision of two heavy nuclei rapidly expands and cools down
until hadronization occur and many aspects of this dynamics are now
fairly well understood in terms of the hydrodynamic modes. This hints
to the existence of a local thermally equilibrated plasma and the
lattice data can accurately describe the thermal QCD contribution
inherent in the experimental data. The question is at what temperature
and chemical potential can one compare the lattice data on
fluctuations of conserved charges to those measured in the
experiments. The traditional method is to determine the temperature
and baryon chemical potential of the fireball at the time of chemical
freeze-out i.e., when all the inelastic scatterings cease to exist
between the hadrons. This involve statistical fits to the ratios of
hadron abundances measured in the experiments \cite{Oeschler:2009im},
inspired from the HRG model, and extracting $T$ and $\mu_B$ such that
the $\chi^2$ per degree of freedom for the fits is closest to unity.
Once these parameters for the fireball are known, it was outlined in
Ref.~\cite{Gavai:2010zn} how lattice QCD data on particular ratios of
fluctuations, proposed in Ref.~\cite{Gupta:2009mu}, can be used to
predict the hadron multiplicities if indeed a thermalized medium is
formed. One of the biggest source of uncertainties is in the
determination of the freeze-out temperature \cite{Gupta:2011wh}. It is
thus important to calculate the freeze-out temperature from first
principles in a model independent way since already several
thermodynamic quantities start to show a departure from the HRG
predictions for $T>140$ MeV. One other issue is to estimate the
curvature of the freeze-out line at finite $\mu_B$ to understand how
far apart in the QCD phase diagram are the chiral and freeze-out
lines.  If the outcome of this study would suggest a close proximity
between these lines, then there is a hope to observe any effects of
the QCD phase transition in the heavy-ion collisions assuming
thermalization occurs.

\subsection{Curvature of the freeze-out line: Input from the lattice}
\begin{figure}
\begin{center}
\includegraphics[width=\textwidth]{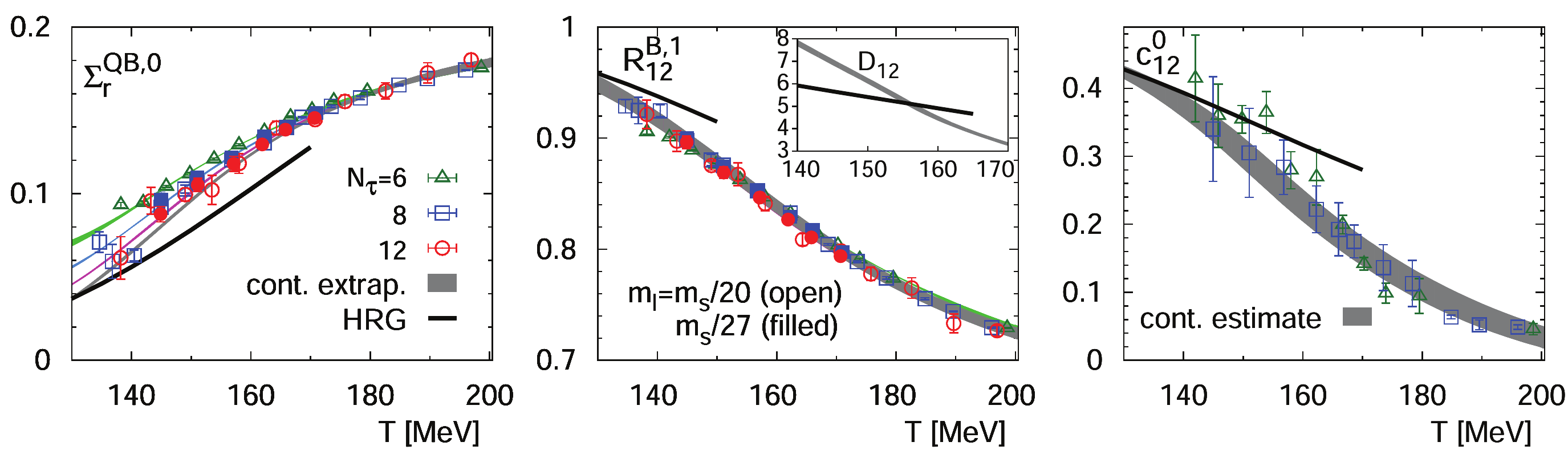}
\end{center}
\caption{The quantities that determine the curvature of freeze-out line show deviation from Hadron Resonance Gas 
predictions already around $145$ MeV, from Ref.~\cite{Bazavov:2015zja}. }
\label{fig:ratios}
\end{figure}

To estimate the curvature of the freeze-out line from first principles
without resorting to hadron resonance gas model inspired fits, a new
method has been proposed by combining lattice and experimental data in
Ref.~\cite{Bazavov:2015zja}.  The freeze-out temperature to the
leading order in baryon chemical potential $\mu_B$ can be parametrized
as $T^f(\mu_B)=T^f(0) \left[1-\kappa_2^f \mu^2_B/T^2\right]$ where
$\kappa_2^f$ is the curvature of the freeze-out curve for small values
of $\mu_B$. The inverse ratio of fluctuations of charges to the total
number of charged species
$R_{12}^Q(T,\mu_B)=\chi_1^Q(T,\mu_B)/\chi_2^Q(T,\mu_B)$ was shown to
act as a ``baryometer'' for the system i.e. at the leading order, give
an estimate of $\mu_B$ of the
fireball~\cite{Bazavov:2012vg}. Equivalently $R_{12}^B$ is also
another suitable candidate for a
``baryometer''~\cite{Bazavov:2012vg}. The ratios of these two simple
quantities in the charge and the baryon sectors respectively,
$R_{12}^{QB}=R_{12}^Q/R_{12}^B$, is recently
proposed~\cite{Bazavov:2015zja} as a suitable observable to extract
the curvature since, in leading order in $\mu_B$
\begin{equation}
 R_{12}^{QB}(\mu_B,T)=R_{12}^{QB}(0,T)\left[R_{12}^{QB,2}-\kappa_2^f ~T_{f,0} \frac{d R_{12}^{QB}(0)}{d T}\vert_{T_{f,0}}
   \right] \frac{\mu_B^2}{T^2}
\end{equation}
In the above expression, the main systematic uncertainty comes from
the uncertainty in estimating the value of $\mu_B$. In order to
measure this ratio in a model independent way under the experimental
conditions assuming thermalization, it could be simply expanded in
terms of the ``baryometer'' $R_{12}^B$. A simple Taylor expansion of
$R_{12}^{QB}(\mu_B,T)$ thus gives,
\begin{equation}
\label{eqn:r12qbmu}
 R_{12}^{QB}(\mu_B)=R_{12}^{QB}(0)\left[1+c_{12}~\left(R_{12}^{B}\right)^2\right]+\mathcal{O}\left(R_{12}^{B}\right)^4~.
\end{equation}
where the curvature of the freeze-out line is then extracted from the relation $c_{12}(T_{f,0},\kappa_2^f)$ $=$ 
$c_{12}(T_{f,0})-\kappa_2^f D_{12}$ with quantities $c_{12},D_{12}$ at $\mu_B=0$ defined as,
\begin{equation}
 c_{12}(T_{f,0})=\frac{1}{(R_{12}^B)^2}\frac{R_{12}^{QB,2}(0)}{R_{12}^{QB}(0)}~,~
 D_{12}(T)=\frac{1}{(R_{12}^B)^2}\frac{d R_{12}^{QB}(0)}{d T}\vert_{T_{f,0}}~.
\end{equation}
The superscript $2$ in all these expression denotes second derivatives of the quantities with respect to $\mu_B$ 
calculated at $\mu_B=0$. The observables at $\mu_B=0$ are very precisely measured on the lattice and even the 
continuum estimates are known~\cite{Bazavov:2015zja}. In order that these ratios are sensitive to the experimental 
conditions, the quantities entering in these ratios at finite values of $\mu_B$ were extracted from the experimental data, 
to provide an estimate of $\kappa_2$ if indeed thermalization is achieved~\cite{Bazavov:2015zja}. There 
are two important caveats to this procedure though. Experimentally, only the net-proton yields are measured and it needs 
to be ascertained how much of the net-baryon number is represented within this measurement. The second 
caveat one has to take into account is that the detectors have a finite momentum acceptance and can only measure hadrons which have transverse momenta beyond some threshold~\cite{Karsch:2015zna}. Surprisingly in the ratio 
$R_{12}^{QB}$, these two systematic uncertainties are argued to almost cancel one another~\cite{Bazavov:2015zja}. 
Within HRG approximation atleast, one can show that the deviation between $R_{12}^{B}$ and for the protons denoted by $R_{12}^P$ 
is within $10\%$ which is equal and opposite in magnitude to the error introduced while neglecting the contribution of the undetected hadrons with momenta $p_T\leq 0.4$ GeV. 
The results for the ratio $R_{12}^{QB}(0)\equiv \Sigma$ at $\mu_B=0$ from Ref.~\cite{Bazavov:2015zja} is summarized in 
Fig.~\ref{fig:ratios}. It is clearly visible that for this observable, the HRG approximation breaks down already at 
$T\sim 145$ MeV. Similar conclusions could be made for the quantities $c_{12}$ and $D_{12}(T)$.

A fit to the experimental data on $R_{12}^{QP}(\mu_B)$ according to Eq. (\ref{eqn:r12qbmu}), yields an intercept $ R_{12}^{QP}(0)$, 
which after comparing with the lattice data on $ R_{12}^{QB}(0)$ gives a freeze-out temperature of $T_f(0)=147(2)$ MeV.
Furthermore from the slope of the fit, the 
curvature comes out to be $\kappa_2^f(0)=-0.001(13)$~\cite{Bazavov:2015zja}. It agrees quite well with the curvature 
of the line of chiral transitions $\kappa_2^B=0.007$ using Taylor expansion method in $\mu_B$~
\cite{Kaczmarek:2011zz,Endrodi:2011gv}, but there is disagreement with the recent results of 
$\kappa_2^B\geq 0.015$ using imaginary chemical potential techniques~\cite{Bonati:2015bha,Bellwied:2015rza,
Cea:2015cya}. 

\begin{figure}
\begin{center}
\includegraphics[width=0.7\textwidth]{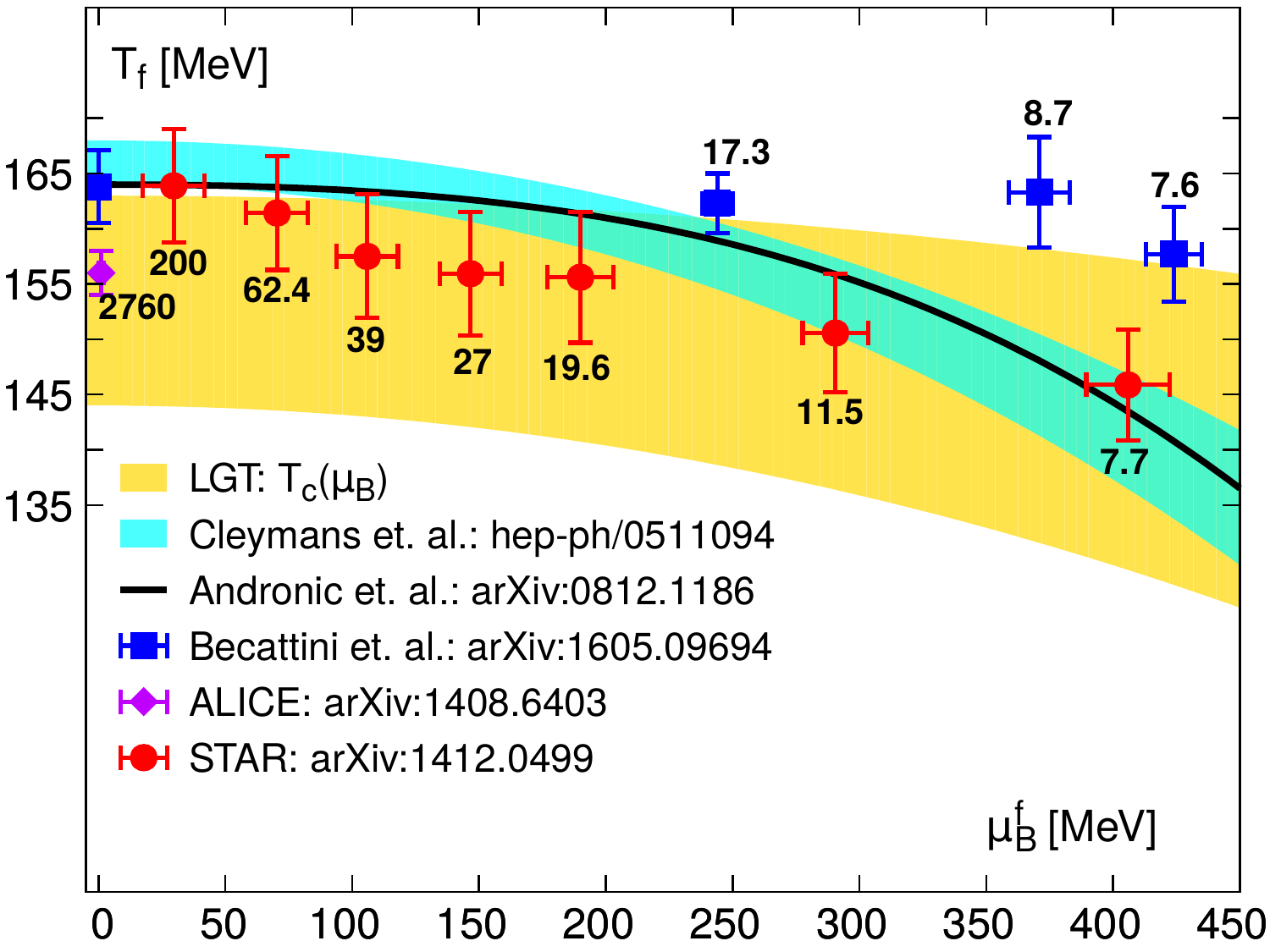}
\end{center}
\caption{The current status of our understanding of heavy-ion experiment data from CERN and RHIC both from 
phenomenological models and from Lattice QCD thermodynamics from Ref.~\cite{Frithjof:cpod}. For more details 
see text.}
\label{fig:curvatureplot}
\end{figure}

The experimental measurements have effects due to the rapidly expanding medium and other systematic biases apart 
from the caveats mentioned earlier.  Surprisingly the general systematic trends of the heavy ion data both from 
RHIC and LHC seem to suggest that these are consistent with predictions from finite temperature QCD using Lattice 
techniques. The current status of our understanding is  outlined succinctly in Fig.~\ref{fig:curvatureplot} taken 
from Ref.~\cite{Frithjof:cpod}.  The yellow band correspond to the current lattice estimate of the chiral curvature 
line and the blue band \cite{Cleymans:2005xv} and the black line \cite{Andronic:2008gu} correspond to different 
statistical model fits to the experimental data. The blue boxes are predictions from a transport model fit to the 
LHC data for different center of mass energies of heavy-ion collisions~\cite{Becattini:2012xb}. It is evident that 
at vanishingly small $\mu_B$, the freeze-out temperatures predicted from lattice are lower than the ones obtained 
from phenomenological freeze-out curves. Though the experimental data points are currently consistent with both lattice 
and phenomenological curves within errors, their central values are systematically lower that the phenomenological 
freeze-out curves. A more precise determination of the freeze-out curve from the lattice would enable to clearly 
distinguish between all these different scenarios. With the new observables and methods like in Ref.~\cite{Bazavov:2015zja} developed
within Lattice QCD thermodynamics will allow us to elucidate the pattern within the complex and rich set of data 
from heavy-ion experiments with an ultimate goal to understand the phase diagram of QCD.

